# Benchmarking (multi)wavelet-based dynamic and static non-uniform grid solvers for flood inundation modelling

Mohammad Kazem Sharifian[1] and Georges Kesserwani[1]

[1]Department of Civil and Structural Engineering, University of Sheffield, Mappin St, Sheffield City Centre, Sheffield S1 3JD, UK.

Corresponding author: Georges Kesserwani (g.kesserwani@sheffield.ac.uk)

**Key Points:**

- Multiwavelet-based solvers are assessed for flood inundation modelling, considering static non-uniform and dynamically adaptive grids

- Adaptive multiwavelet second-order discontinuous Galerkin (MWDG2) solver outperforms for modelling rapidly propagating flows

- Non-uniform DG2 and local inertial ACC solvers are more accurate and efficient choices when modelling slow-to-gradually propagating flows

**Abstract**

This paper explores static non-uniform grid solvers that adapt three raster-based flood models on an optimised non-uniform grid: the second-order discontinuous Galerkin (DG2) model representing the modelled data as piecewise-planar fields, the first-order finite volume (FV1) model using piecewise-constant fields, and the local inertial (ACC) model only evolving piecewise-constant water depth fields. The optimised grid is generated by applying the multiresolution analysis (MRA) of multiwavelets (MWs) to piecewise-planar representation of raster-formatted topography data, for more sensible grid coarsening based on one user-specified parameter. Two adaptive solvers are also explored that apply the MRA of MWs and of Haar wavelets (HWs) to, respectively, scale and adapt the DG2 (MWDG2) and FV1 (HWFV1) modelled data dynamically in time. The performance of the non-uniform grid and adaptive solvers is assessed in terms of flood depth and extent, velocities, and CPU runtimes, with reference to the raster-based DG2 model predictions on their finest resolution grid. The assessments considered three large-scale flooding scenarios, involving rapid and slow-to-gradual flows. MWDG2 is found to be the most favourable choice when modelling rapid flows, where it excels in capturing small velocity variations. For slow-to-gradual flows, the adaptive solvers deliver less accurate outcomes, and their efficiency can be hampered by overhead costs of the dynamic MRA. Instead, non-uniform DG2 is recommended to capture urban flow interactions more accurately. Non-uniform ACC is 5 times faster to run than non-uniform DG2 but delivers close flooding depth and extent predictions, thus is more attractive for fluvial/pluvial flood simulation over large areas.

# 1. Introduction

Flood modelling is an essential tool to understand flooding dynamics and define mitigation strategies to support decision-making in flood risk management for reducing the potential loss of life and property (Spiekermann et al., 2015). With the advances made in remote sensing technologies and the increased availability of high-quality terrain data, two-dimensional (2D) raster-based hydrodynamic models have gained popularity for large-scale flood modelling (Hunter et al., 2008; Teng et al., 2017; Nkwunonwo et al., 2020). These models are based on solving a form of shallow water equations (SWE) using a numerical solver on uniform resolution grids defined by a raster-formatted digital elevation model (DEM), which leverage local and compact computational stencil. A variety of models exist; some are based on applying simplifications in the SWE, such as the zero-inertia approximation (Aricò et al., 2011; Leandro et al., 2014) and the local inertial approximation such as the ACC solver (Bates et al., 2010; Neal et al., 2018); some others have adopted the numerical solution of the full SWE using the Godunov-type framework within first-order finite volume (FV1) (Echeverribar et al., 2019; Sanders & Schubert, 2019) and second-order discontinuous Galerkin (DG2) solvers (Shaw et al., 2021; Ayog et al., 2021).

As the flood flow patterns on large areas are strongly influenced by topographic features or sharp-edged man-made structures (e.g., buildings, roads, defences, curbs on the urban surface), fine grid resolution is often required to produce reliable flood maps and hydrographs (Yu & Harbor, 2019; Bellos et al., 2020; Muthusamy et al., 2021; Xing et al., 2021). Raster-based solvers on uniform resolution grids can therefore lead to increased computational cost as the spatial extent of the area becomes larger and when simulating over a long duration (Bellos et al., 2020; Saksena et al., 2020; Guo et al., 2020). This has motivated several lines of efforts to overcome the computational cost barrier, from the use of subgrid (Sanders et al., 2008; Casulli & Stelling, 2011; Özgen et al., 2016) and local time step methods (Sanders, 2008; Kesserwani & Liang, 2015; Dazzi et al., 2018; van den Bout & Jetten, 2020) to parallel processing (Vacondio et al., 2017; Xia et al., 2019; Morales-Hernández et al., 2021). From another perspective, Shaw et al. (2021) demonstrated that the smoothness of the piecewise-planar basis of the DG2 solver in representing flow variables and topography allows improved modelling accuracy over the piecewise-constant basis of the FV1 and ACC solvers on 2-10 fold coarser resolutions. Still, the reliability of the DG2 prediction requires the river channel to be wider than half of the grid resolution. As capturing the small-scale topographic features is central to predict the spatial and temporal scales of the flood flow response and propagation over large areas, one option to alleviate the computational burden of raster-based solvers is to deploy multiresolution alternatives that apply the flow solvers on a non-uniform grid. An efficiently generated non-uniform grid is expected to concentrate the finest resolutions of the DEM around the key topographic drivers of the flood, while retaining as coarse resolution as possible in the other portions of the grid. A smart approach is therefore desired to generate such an optimised non-uniform grid, in a cost-effective way without introducing extra burdens on the user as is often the case with conventional adaptive mesh refinement (AMR) methods (e.g., Popinet, 2011; Zhou et al., 2013; Mandli & Dawson, 2014).

The approach to generate a non-uniform grid using wavelet-based multiresolution analysis (MRA) allows to overcome many inconveniences of conventional AMR methods, such as reducing the user input parameters to one and deploying rigorous formulas to upscale, downscale and recover the modelled data across a cascade of resolution scales (Gerhard & Müller, 2016). The MRA can be naturally fed into the basis of Godunov-type uniform grid solvers to allow them to automatically analyse, scale and assemble flow and topography data onto non-uniform data and grid (Gerhard et al., 2015). Such a form of dynamically adaptive solvers preserves the predictive quality of the uniform grid solvers, and their non-uniform grids are not necessarily restricted by the grading procedure, or 2:1 rule, that is applied in conventional AMR methods, (Liang, 2011; Caviedes-Voullième et al., 2020). Kesserwani and Sharifian (2020) formulated multiwavelet second-order discontinuous Galerkin (MWDG2) and Haar wavelet first-order finite volume (HWFV1) solvers that robustly handle the presence of wet-dry fronts across steep topographic slopes, with validations for shallow water flow benchmarks including laboratory-scale dam-break and tsunami scenarios. Their findings particularly revealed a great potential for the MWDG2 solver to retrieve the modelling quality of the uniform DG2 solver on the finest resolution grid at a runtime cost that is competitive with that of the uniform FV1 solver. Moreover, the MRA of MWs was found to lead to more sensible scaling of the modelled data than HWs, pointing to a promising potential to further exploit MWs to generate optimised non-uniform grids, smartly and cost-effectively from highest resolution DEM available. So far, the adaptive MWDG2 and HWFV1 solvers were only assessed for academic test cases involving rapidly propagating flows, awaiting to be further benchmarked for more realistic and large-scale flooding scenarios. On the other hand, the use of MRA to generate static non-uniform grids has been limited to HWs: Özgen-Xian et al. (2020) used the MRA of HWs to generate unstructured triangular meshes on which a zero-inertia solver was adapted to simulate large-scale flood flows. They reported 2-3 fold speed-ups against the uniform grid counterpart and concluded that the accuracy of their non-uniform grid solver simulations is dependent on both the smoothness and resolution of the topography data. These moderate speed-ups or accuracy issues are likely to be improved by optimised non-uniform grids, generated using the MRA of MWs, on which the more complex ACC, FV1 and DG2 flood flow solvers are adapted as non-uniform solvers.

This work has a twofold aim. First, to explore the potential of using the optimised non-uniform grids generated by the MRA of MWs to reduce the cost of the raster-based ACC, FV1 and DG2 flow solvers; and second, to further study the performance of the adaptive MWDG2 and HWFV1 solvers alongside the non-uniform ACC, FV1 and DG2 solvers for large-scale flood modelling applications. The paper is therefore structured as follows: Section 2 starts with an overview of the raster-based ACC, FV1 and DG2 solvers, then describes the optimised non-uniform grid generation procedure from the MRA of MWs applied to scale and analyse piecewise-planar representations of the raw raster-formatted DEM data. This section also describes how the solvers are applied on the optimised grid to get non-uniform ACC, FV1 and DG2 solver counterparts, and overviews how the MRA is applied within the adaptive MWDG2 and HWFV1 solvers. In Section 3, the adaptive and non-uniform solvers are benchmarked for three large-scale flood modelling case studies to assess their predictive capability and efficiency compared to reference predictions from the raster-based DG2 solver on the finest uniform resolution available. The first test case simulates an Environment Agency benchmark (Néelz & Pender, 2013) involving a flash flood along a narrow valley, to analyse the response of models in reproducing rapidly and gradually propagating flows. In the second test case, the solvers are assessed in simulating a hypothetical flood induced by a defence failure in the Thames Barrier (Liang et al., 2008), focusing on their capabilities in modelling gradually propagating flows over low-lying areas with complex terrain features. The final test case reproduces a flood event that occurred in Carlisle in 2005 (Neal et al., 2009) to explore the solvers' performance when modelling real-world flooding scenarios over realistic urban topography driven by multiple riverine inflows. Conclusions are finally drawn in Section 4.

**2. Methods**

This section summarises the implementation of the non-uniform DG2, FV1 and ACC solvers on an optimised grid generated by the MRA of MWs. Section 2.1 overviews the technical aspects of the raster-based uniform DG2, FV1 and ACC solvers. Section 2.2 describes how the MRA of MWs is applied to the piecewise-planar representation for the topography data from a raw raster-formatted DEM to generate an optimised static non-uniform grid, on which the raster-based DG2, FV1 and ACC solvers are adapted, including the treatments applied to evaluate the flow data at the interfaces separating non-homogeneous elements with different resolutions. Section 2.3 outlines the adaptive MWDG2 and HWFV1 solvers that are also included in the comparisons in Section 3.

**2.1. Raster-based shallow water flow solvers**

The raster-based uniform DG2, FV1 and ACC solvers are fully detailed in Shaw et al. (2021), with a link to access their source codes. They are briefly overviewed to contrast the difference in their mathematical and numerical complexity and show how their modelled data can be expressed in terms of a scaled local basis function $\phi$ from which the MRA of (multi)wavelets is compatibly derived.

These solvers are formulated to simulate flooding over a 2D domain of dimension $X \times Y$. The finest raster grid of raw topography data has a uniform resolution of $R = \Delta x = \Delta y$, and the domain is subdivided into non-overlapping square elements indexed by $i, j$ and centred at points $(x_{i,j}, y_{i,j})$, as shown in Figure 1.

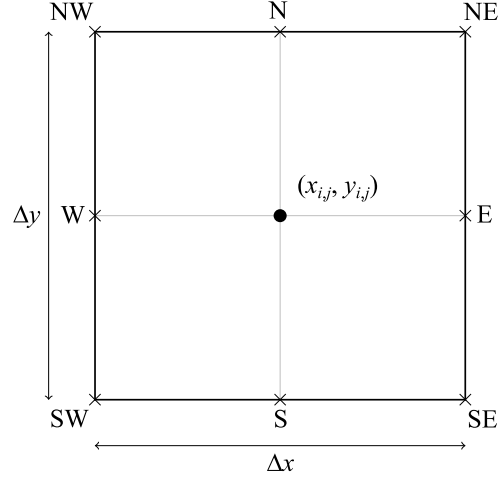

**Figure 1.** A square element centred at $(x_{i,j}, y_{i,j})$ with horizontal dimension $R = \Delta x = \Delta y$, on which raster-based DG2, FV1 and ACC solvers are applied. N, E, S and W mark the northern, eastern, southern and western interface centres, and NW, NE, SW and SE mark the four vertices.

**2.1.1. The DG2 solver**

The DG2 solver has the highest level of complexity both mathematically and numerically. It solves the conservative vectorial form of the full SWE,

$$\partial_t \mathbf{U} + \partial_x \mathbf{F}(\mathbf{U}) + \partial_y \mathbf{G}(\mathbf{U}) = \mathbf{S}_b(\mathbf{U}) + \mathbf{S}_f(\mathbf{U}) \qquad (1)$$

where $\partial$ represents a partial derivative operator, $\mathbf{U}(x, y, t) = [h, q_x, q_y]^{\mathrm{T}}$ is the vector of flow variables at time $t$ and location $(x, y)$ which contains the water depth $h$ (m) and the discharges per unit width, $q_x = hu$ (m²/s) and $q_y = hv$ (m²/s), involving the depth-averaged horizontal velocities $u$ (m/s) and $v$ (m/s) in the Cartesian directions, respectively; $\mathbf{F} = [q_x, q_x^2/h + gh^2/2, q_x q_y/h]^{\mathrm{T}}$ and $\mathbf{G} = [q_y, q_x q_y/h, q_y^2/h + gh^2/2]^{\mathrm{T}}$ are vectors including the components of physical flux field, with $g$ (m/s²) denoting the gravity acceleration constant. $\mathbf{S}_b = [0, -gh\partial_x z, -gh\partial_y z]^{\mathrm{T}}$ is the source term vector including the topography gradients and $\mathbf{S}_f = [0, -C_f u\sqrt{u^2 + v^2}, -C_f v\sqrt{u^2 + v^2}]^{\mathrm{T}}$ is the source term vector incorporating friction effects expressed as function of $C_f = g n_M^2 / h^{1/3}$ where $n_M$ (s/m$^{1/3}$) refers to the Manning resistance parameter.

In solving Equation 1, the discrete vector of flow variables $\mathbf{U}_h(x, y, t)$ and scalar topography $z_h(x, y)$ are shaped as local piecewise-planar solutions over each element, by adopting a simplified and scaled Legendre basis $\boldsymbol{\phi}$ (Kesserwani et al., 2018; Kesserwani & Sharifian, 2020). $\mathbf{U}_h(x, y, t)$ is therefore spanned by three coefficients, an average (denoted by subscript 0) and two slopes in the $x$ and $y$ directions (denoted by subscripts $1x$ and $1y$, respectively). With these coefficients, each physical component of $\mathbf{U}_h(x, y, t)$ is represented via an equation of a plane over each element as,

$$\mathbf{U}_h(x, y, t) = \mathbf{U}_{i,j} \cdot \boldsymbol{\phi} = \begin{bmatrix} h_{i,j,0} & h_{i,j,1x} & h_{i,j,1y} \\ {q_x}_{i,j,0} & {q_x}_{i,j,1x} & {q_x}_{i,j,1y} \\ {q_y}_{i,j,0} & {q_y}_{i,j,1x} & {q_y}_{i,j,1y} \end{bmatrix} \cdot \begin{bmatrix} 1 \\ 2\sqrt{3}(x - x_{i,j})/R \\ 2\sqrt{3}(y - y_{i,j})/R \end{bmatrix} \quad (2)$$

where $\mathbf{U}_{i,j}$ is a matrix of flow coefficients where each of its rows include the average and slope coefficients defining the piecewise-planar representation of each component of the flow vector after projection onto the basis $\boldsymbol{\phi}$ (note that the "·" in Equation 2 stands for scalar product). Similarly, $z_h(x, y)$ is locally expanded by three coefficients as follows,

$$z_h(x, y) = \mathbf{z}_{i,j} \cdot \boldsymbol{\phi} = \begin{bmatrix} z_{i,j,0} & z_{i,j,1x} & z_{i,j,1y} \end{bmatrix} \cdot \begin{bmatrix} 1 \\ 2\sqrt{3}(x - x_{i,j})/R \\ 2\sqrt{3}(y - y_{i,j})/R \end{bmatrix} \quad (3)$$

where $\mathbf{z}_{i,j}$ is a vector, including the average and slope coefficients of topography, for defining local piecewise-planar representations of the topography after projection onto the same basis $\boldsymbol{\phi}$ (Shaw et al., 2021). The time-invariant topography coefficients $\mathbf{z}_{i,j}$ are projected from the raw DEM raster file, as described later in Sect. 2.1.4. The time-dependent flow coefficients $\mathbf{U}_{i,j}$ are initialised component-wise from a given initial condition defined through a raster file, in the same way.

To evolve the initialised flow coefficients in matrix $\mathbf{U}_{i,j}$ from time level $n$ to $n + 1$, Equation 1 is discretised by an explicit second-order two-stage Runge-Kutta scheme (Kesserwani et al., 2018),

$$\mathbf{U}_{i,j}^{\text{int}} = \mathbf{U}_{i,j}^n + \Delta t \, \mathbf{L}(\mathbf{U}_{i,j}^n) \tag{4a}$$

$$\mathbf{U}_{i,j}^{n+1} = \frac{1}{2}\left[\mathbf{U}_{i,j}^n + \mathbf{U}_{i,j}^{\text{int}} + \Delta t \, \mathbf{L}(\mathbf{U}_{i,j}^{\text{int}})\right] \tag{4b}$$

where the time step $\Delta t$ is calculated according to the CFL condition using the maximum stable Courant number of 0.33 (Cockburn & Shu, 2001) and $\mathbf{L} = \left[\mathbf{L}_0, \mathbf{L}_{1x}, \mathbf{L}_{1y}\right]$ is a 3×3 matrix where each of its columns include the components of discrete spatial operator vectors, $\mathbf{L}_0$, $\mathbf{L}_{1x}$ and $\mathbf{L}_{1y}$. These local operators contain the spatial fluxes and topography gradient terms (Shaw et al., 2021), but after applying temporary revisions to the components of $\mathbf{U}_{i,j}$ to ensure well-balancedness of the flow over uneven topography and non-negative water depths in the presence of wet-dry fronts (Kesserwani & Sharifian, 2020). The fluxes across the four interfaces of each element are computed using the HLL approximate Riemann solver based on the revised limits of the piecewise-planar solutions of the flow vector at the two sides of each interface (Figure 1). The friction source term $\mathbf{S}_f$ is not explicitly included in $\mathbf{L}$ but instead discretised using a split implicit scheme applied at the beginning of each time step.

### 2.1.2. The FV1 solver

The FV1 scheme solves the same conservative form of the full SWE in Equation 1, as the DG2, but simplifies numerical complexity by representing $\mathbf{U}_h(x,y,t)$ and $z_h(x,y)$ as piecewise-constant solutions per element, by only considering the first component of the basis $\boldsymbol{\phi}$ or setting it to be equal to 1. By doing so, each physical component of $\mathbf{U}_h(x,y,t)$ takes the form of a flat plane that is only defined by an average coefficient as,

$$\mathbf{U}_h(x,y,t) = \mathbf{U}_{i,j} \cdot \boldsymbol{\phi} = \begin{bmatrix} h_{i,j,0} \\ q_{x\,i,j,0} \\ q_{y\,i,j,0} \end{bmatrix} \cdot 1 = \begin{bmatrix} h_{i,j,0} \\ q_{x\,i,j,0} \\ q_{y\,i,j,0} \end{bmatrix} \tag{5}$$

Here, $\mathbf{U}_{i,j}$ is a vector of average flow coefficients and $z_h(x,y)$ is represented in the same way as,

$$z_h(x,y) = \mathbf{z}_{i,j} \cdot \boldsymbol{\phi} = z_{i,j,0} \cdot 1 = z_{i,j,0} \tag{6}$$

For the FV1 spatial discretisation, the matrix **L** only considers the $\mathbf{L}_0$ vector, while the HLL approximate Riemann solver takes in the piecewise-constant solution limits at the two sides of each interface. Here, the vector of average coefficient $\mathbf{U}_{i,j}$ is evolved from time level $n$ to $n+1$ using a forward Euler time-stepping scheme as follows,

$$\mathbf{U}_{i,j}^{n+1} = \mathbf{U}_{i,j}^{n} + \Delta t \, \mathbf{L}(\mathbf{U}_{i,j}^{n}) \tag{7}$$

and the maximum stable Courant number of 0.5 is used to calculate the time step $\Delta t$ (Kesserwani & Liang, 2012b).

### 2.1.3. The ACC solver

The ACC solver uses the local inertial equations that neglect the acceleration term of the full SWE (Bates et al., 2010). It mainly stores averaged water depth and topography data as piecewise-constant elementwise, as with the FV1 solver. The water depth is evolved in time using the same philosophy as the FV1, however based on a finite difference computation of continuous discharges across four interfaces of each element (Figure 1), instead of using a Riemann solver (De Almeida et al., 2012). These discharges are estimated based on coupling the simplified momentum conservation equation to the Manning resistance formula (De Almeida et al., 2012). The maximum stable Courant number of 0.7 has been used to calculate the time step with the ACC solver.

### 2.1.4. Initialisation of piecewise-planar topography and flow coefficients

The coefficients $\mathbf{z}_{i,j} = [z_{i,j,0}, z_{i,j,1x}, z_{i,j,1y}]$ are projected on the uniform grid so that $z_h(x,y)$ remains continuous at interface centres of an element (denoted by N, S, E and W, in Figure 1) where Riemann fluxes are evaluated from the well-balancedness and positivity-preserving revisions of the limits of the local solutions (Kesserwani et al., 2018). The topography is estimated at these interface centres by averaging the DEM raster values taken at the NW, NE, SW and SE vertices, as $z_{i,j}^{N} = (z_{i,j}^{NW} + z_{i,j}^{NE})/2$ and in the same manner for $z_{i,j}^{E}$, $z_{i,j}^{S}$ and $z_{i,j}^{W}$. From these estimates, the average $z_{i,j,0}$ and slope coefficients $z_{i,j,1x}$ and $z_{i,j,1y}$ can be defined as (Shaw et al., 2021),

$$z_{i,j,0} = \frac{1}{2}[z_{i,j}^{\text{E}} + z_{i,j}^{\text{W}}] = \frac{1}{2}[z_{i,j}^{\text{N}} + z_{i,j}^{\text{S}}] = \frac{1}{2}[z_{i,j}^{\text{NE}} + z_{i,j}^{\text{NW}} + z_{i,j}^{\text{SE}} + z_{i,j}^{\text{SW}}] \tag{8a}$$

$$z_{i,j,1x} = \frac{1}{2\sqrt{3}}[z_{i,j}^{\text{E}} - z_{i,j}^{\text{W}}] = \frac{1}{4\sqrt{3}}[z_{i,j}^{\text{NE}} - z_{i,j}^{\text{NW}} + z_{i,j}^{\text{SE}} - z_{i,j}^{\text{SW}}] \tag{8b}$$

$$z_{i,j,1y} = \frac{1}{2\sqrt{3}}[z_{i,j}^{\text{N}} - z_{i,j}^{\text{S}}] = \frac{1}{4\sqrt{3}}[z_{i,j}^{\text{NE}} - z_{i,j}^{\text{SE}} + z_{i,j}^{\text{NW}} - z_{i,j}^{\text{SW}}] \tag{8c}$$

The coefficients $[h_{i,j,0}, h_{i,j,1x}, h_{i,j,1y}]$, $[q_{x_{i,j,0}}, q_{x_{i,j,1x}}, q_{x_{i,j,1y}}]$ and $[q_{y_{i,j,0}}, q_{y_{i,j,1x}}, q_{y_{i,j,1y}}]$ of the physical components of the flow vector $\mathbf{U}_h(x, y, t)$ are initialised in the same manner as Equation 8.

**2.2. Multiwavelet-based static non-uniform grid generation**

Assuming that the finest resolution $R$ on the uniform grid is associated with a maximum resolution level $L$, this grid would be composed of $M \times 2^L$ and $N \times 2^L$ elements in the $x$ and $y$ directions where $M = X/2^L R$ and $N = Y/2^L R$. The MRA of MWs starts with the topography coefficients at level $L$, $\mathbf{z}^{\text{fine}}$ to encode, or produce, topography coefficients at a twice-coarser resolution grid at level $L - 1$, $\mathbf{z}^{\text{coarse}}$, and three vectors of details $\mathbf{d}_{\text{H}}^{\text{coarse}}$, $\mathbf{d}_{\text{V}}^{\text{coarse}}$ and $\mathbf{d}_{\text{D}}^{\text{coarse}}$. These vectors encapsulate the encoded differences in topography coefficients between the two resolution levels $L$ and $L - 1$ along the horizontal (H), vertical (V) and diagonal (D) directions, respectively (Kesserwani & Sharifian, 2020). The encoding process is applied as follows:

$$\mathbf{z}^{\text{coarse}} = \mathbf{HH}^0 \mathbf{z}^{\text{fine}}_{[0]} + \mathbf{HH}^1 \mathbf{z}^{\text{fine}}_{[2]} + \mathbf{HH}^2 \mathbf{z}^{\text{fine}}_{[1]} + \mathbf{HH}^3 \mathbf{z}^{\text{fine}}_{[3]} \tag{9a}$$

$$\mathbf{d}_{\text{H}}^{\text{coarse}} = \mathbf{GA}^0 \mathbf{z}^{\text{fine}}_{[0]} + \mathbf{GA}^1 \mathbf{z}^{\text{fine}}_{[2]} + \mathbf{GA}^2 \mathbf{z}^{\text{fine}}_{[1]} + \mathbf{GA}^3 \mathbf{z}^{\text{fine}}_{[3]} \tag{9b}$$

$$\mathbf{d}_{\text{V}}^{\text{coarse}} = \mathbf{GB}^0 \mathbf{z}^{\text{fine}}_{[0]} + \mathbf{GB}^1 \mathbf{z}^{\text{fine}}_{[2]} + \mathbf{GB}^2 \mathbf{z}^{\text{fine}}_{[1]} + \mathbf{GB}^3 \mathbf{z}^{\text{fine}}_{[3]} \tag{9c}$$

$$\mathbf{d}_{\text{D}}^{\text{coarse}} = \mathbf{GC}^0 \mathbf{z}^{\text{fine}}_{[0]} + \mathbf{GC}^1 \mathbf{z}^{\text{fine}}_{[2]} + \mathbf{GC}^2 \mathbf{z}^{\text{fine}}_{[1]} + \mathbf{GC}^3 \mathbf{z}^{\text{fine}}_{[3]} \tag{9d}$$

where $\mathbf{HH}^{0,1,2,3}$ are 3×3 low-pass filter matrices and $\mathbf{GA}^{0,1,2,3}$, $\mathbf{GB}^{0,1,2,3}$ and $\mathbf{GC}^{0,1,2,3}$ are 3×3 high-pass filter matrices that were generated from the MWs decomposition of the basis $\boldsymbol{\phi}$ (Kesserwani & Sharifian, 2020). The subscripts of vectors $\mathbf{z}^{\text{fine}}$ refer to the numbering of the four child elements at level $L$, with reference to their parent element at level $L-1$. By recursive application of Equation 9, $L$ times, the topography coefficients at level $L$ can equivalently be represented by a series of details, that sum up over the coarsest resolution topography coefficients at level 0 (i.e., on the coarsest grid made of $M \times N$ elements). On a hierarchy of grids with resolutions levels varying between 0 and $L$, these details become increasingly significant with increasing non-smoothness in the topography features while remain negligible in the other areas where the topography is smooth.

To identify the areas that require high resolution, the significance of the details is therefore analysed by comparing their magnitude to an error threshold parameter $\varepsilon$. The parameter $\varepsilon$ is usually specified by the user and a value of $10^{-3}$ is recommended for shallow flow modelling to keep a sensible balance between efficiency and accuracy (Kesserwani et al., 2019; Kesserwani & Sharifian, 2020). To make sure that all details are analysed in a problem-independent manner, they must be normalised. A normalised detail is computed in each element as $\check{d} = |d|/\max(1, |z_0|)$, where $|d|$ is the maximum of all the topography details along horizontal, vertical and diagonal directions, and $|z_0|$ denotes the maximum of all average coefficients of topography on the uniform grid at level $L$. The normalised details are computed for all elements with resolution levels $L-1, \ldots, 0$, while flagging those elements with significant details for refinement, i.e., to preserve their details (Kesserwani & Sharifian, 2020). To ensure significant details can be re-accessed in a tree-like hierarchy, if a child element is flagged to have a significant detail, all its parent elements on coarser levels will be flagged as well.

Once all elements with significant details are identified, the topography coefficients on the hierarchy of grids can be assembled to produce a non-uniform grid with various resolution levels. The assembly process is performed by adding up the remaining significant details to the coarsest topography coefficients via decoding. The decoding starts from level 0, using Equation 10, to produce the topography coefficients at four child elements at level 1, $\mathbf{z}_{[0]}^{\text{fine}}$, $\mathbf{z}_{[1]}^{\text{fine}}$, $\mathbf{z}_{[2]}^{\text{fine}}$ and $\mathbf{z}_{[3]}^{\text{fine}}$, from the topography coefficients of the parent element at level 0, $\mathbf{z}^{\text{coarse}}$, and the encoded details $\mathbf{d}_{\text{H}}^{\text{coarse}}$, $\mathbf{d}_{\text{V}}^{\text{coarse}}$ and $\mathbf{d}_{\text{D}}^{\text{coarse}}$ (Kesserwani & Sharifian, 2020).

$$\mathbf{z}_{[0]}^{\text{fine}} = [\mathbf{HH}^0]^T \mathbf{z}^{\text{coarse}} + [\mathbf{GA}^0]^T \mathbf{d}_H^{\text{coarse}} + [\mathbf{GB}^0]^T \mathbf{d}_V^{\text{coarse}} + [\mathbf{GC}^0]^T \mathbf{d}_D^{\text{coarse}} \quad (10a)$$

$$\mathbf{z}_{[2]}^{\text{fine}} = [\mathbf{HH}^1]^T \mathbf{z}^{\text{coarse}} + [\mathbf{GA}^1]^T \mathbf{d}_H^{\text{coarse}} + [\mathbf{GB}^1]^T \mathbf{d}_V^{\text{coarse}} + [\mathbf{GC}^1]^T \mathbf{d}_D^{\text{coarse}} \quad (10b)$$

$$\mathbf{z}_{[1]}^{\text{fine}} = [\mathbf{HH}^2]^T \mathbf{z}^{\text{coarse}} + [\mathbf{GA}^2]^T \mathbf{d}_H^{\text{coarse}} + [\mathbf{GB}^2]^T \mathbf{d}_V^{\text{coarse}} + [\mathbf{GC}^2]^T \mathbf{d}_D^{\text{coarse}} \quad (10c)$$

$$\mathbf{z}_{[3]}^{\text{fine}} = [\mathbf{HH}^3]^T \mathbf{z}^{\text{coarse}} + [\mathbf{GA}^3]^T \mathbf{d}_H^{\text{coarse}} + [\mathbf{GB}^3]^T \mathbf{d}_V^{\text{coarse}} + [\mathbf{GC}^3]^T \mathbf{d}_D^{\text{coarse}} \quad (10d)$$

Equation 10 is recursively applied in ascending order to move over the tree of details, while inspecting the details and subsequently creating child elements. The decoding is stopped when reaching an element with an insignificant detail, and finally creates assembled topography coefficients on an optimised non-uniform grid made of non-overlapping elements with different resolutions varying between $R$ and $2^L R$.

**2.2.1. Evaluating the local solutions at non-homogeneous interfaces**

The grid generation framework described in Section 2.2 results in a non-uniform grid with arbitrary resolutions, leading to adjacent elements with different sizes, hence with non-homogeneous interfaces between them to process modelled data that are more than one resolution level apart from each side of these interfaces. An example is shown in Figure 2a, where the red elements are two levels finer than the neighbouring blue element. With the ACC solver, the computational stencil stores and evolves the water depth and topography at the element centres from single-valued discharge estimates at the interfaces, which makes it directly adaptable on the non-uniform grid: the discharges through the finer interfaces (denoted by yellow and red points) can be computed in the same manner as the uniform grid, and then aggregated to give the discharge through the coarse interface (denoted by the blue point) to be used to update the water depth in the coarse element.

With the FV1 and DG2 solvers, the Riemann fluxes across the finer interfaces can also be computed by direct evaluation of the local solution limits at the two sides of the interface centres. However, at the right side of the coarse interface centre there is no straightforward way to evaluate the local solution limit, which is required for computing the revised well-balanced and positivity-preserving flow coefficients from the side of the coarse element (Kesserwani & Sharifian, 2020).

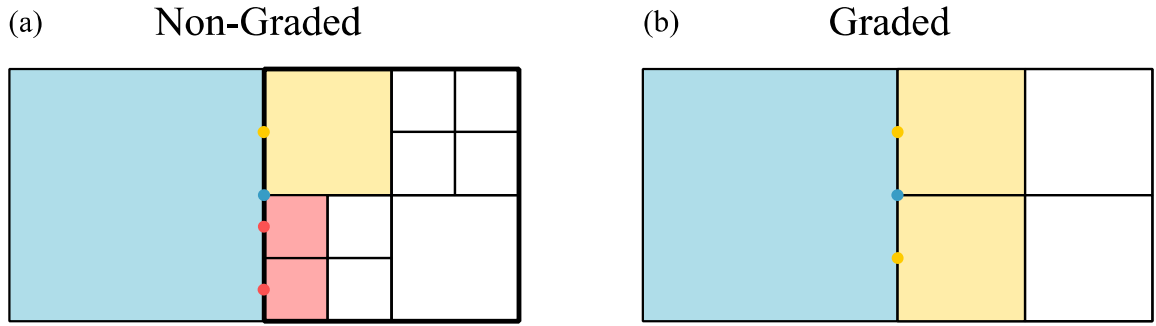

**Figure 2.** Schematic layout of non-homogeneous interfaces between neighbouring elements at different levels: (a) non-graded grid, where a coarse element (in blue) is adjacent to one- and two-level finer elements (in yellow and red, respectively); (b) graded grid, where a coarse element (in blue) can only be adjacent to one-level finer elements (in yellow). The blue, yellow and red points denote the interface centres where the limits of the local solutions are required for the Riemann solvers. The thick border denotes the locally coarsened adjacent element (see Section 2.3).

A common treatment that most conventional AMR methods deploy to alleviate this issue is to grade or regularise the grid by ensuring that no element is more than one level finer or coarser than its neighbouring elements (Borthwick et al., 2000; Popinet, 2003; Kesserwani & Liang, 2012a; Donat et al., 2014; Ghazizadeh et al., 2020; Dunning et al., 2020), as shown in Figure 2b. By applying the grading constraint, the required local solution limit at the right side of the coarse interface can be interpolated from the solution limits at the two finer interface centres, denoted by yellow points in Figure 2b. This interpolation is exact with the FV1 solver as its local solution representation is piecewise-constant on each element (i.e., takes the same value along the non-homogenous interfaces), but it is expected to be approximate for the DG2 solver due to the piecewise-planar representation of the local solution (i.e., takes different values along the non-homogenous interfaces as the flow becomes more dynamic). It should be noted that for the sake of consistency in the comparisons among the solvers, the ACC solver has also been applied on the graded non-uniform grid.

**2.3. (Multi)wavelet-based dynamically adaptive solvers**

The adaptive MWDG2 and HWFV1 solvers introduced in Kesserwani and Sharifian (2020) are built by following the same algorithm used to generate static non-uniform grids (Section 2.2). However,

MWDG2 and HWFV1 perform the MRA dynamically at each time step, by considering both flow and topography data. In the adaptive MWDG2 solver, the flow and topography coefficients, $\mathbf{U}_{i,j}$ and $\mathbf{z}_{i,j}$, on the uniform grid at resolution level $L$ are recursively encoded using Equation 9 to get the flow and topography details over levels $0, \ldots, L-1$. The normalised details are then computed and measured against the same $\varepsilon$, to form a tree of significant details. The flow and topography details are recursively decoded over this tree to assemble the flow and topography coefficients on a non-uniform grid on which the DG2 solver is applied to update the matrix of flow coefficients. The updated coefficients are again subjected to encoding to repeat the process dynamically until the end of the simulation. The HWFV1 solver follows the same process as the MWDG2, with the difference of using MRA of HW's to dynamically scale the piecewise-constant flow and topography data, and thereby the grid resolution over time.

In adaptive solvers, the dynamic scaling of modelled data provides a natural means to rigorously evaluate the solution limits at both sides of the non-homogeneous interfaces without needing any heuristic interpolation (Section 2.2.1). Namely, local encoding of the coefficients on the set of finer elements can be applied to compute their exact coefficients required at the coarser resolution that are well-balanced and positivity-preserving (Kesserwani & Sharifian, 2020). Note that applying encoding for the static non-uniform grid solvers is not recommended: as these solvers do not scale their modelled flow data dynamically, encoding can lead to the accumulation of adaptation errors at the same location that can bias the overall model outcomes (Kesserwani et al., 2019).

**3. Results and discussions**

The adaptive and non-uniform grid solvers described in Section 2 are used to reproduce three large-scale test cases. The first test is an industry-standard benchmark of a flash flood, generated from a hypothetical dam-break event, propagating inside a valley with rugged terrain (Section 3.1). It is aimed to study the response of the solvers for both rapidly and gradually propagating flood types. The second test has been commonly used as a benchmark to study adaptive grid refinement strategies with flood simulators; it involves a gradually propagating flood caused by hypothetical defence failure in the Thames Barrier (Section 3.2). The third test reproduces the 2005 fluvial flood event that occurred in Carlisle (Section 3.3), for which the simulations cover the scale of an entire city while including complex urban features and flooding from multiple riverine inflows.

For each of the test cases, a DG2 simulation is run on the finest uniform grid to produce a reference solution against which the adaptive and non-uniform solvers are benchmarked. Runs using the adaptive MWDG2 and HWFV1 solvers are configured such that their maximum level $L$ is at a resolution matching the finest uniform grid, while the non-uniform grid solvers are run on the grids generated by the MRA of MWs at $\varepsilon = 10^{-3}$ and after grading. All the runs were conducted on the same desktop computer using a single-core CPU for a consistent comparison, and the simulation results are openly available on Zenodo (Sharifian & Kesserwani, 2021).

To allow for quantitative comparisons between the outputs produced by the adaptive and non-uniform solvers and the reference uniform DG2 solutions, recommended metrics are used (Nguyen et al., 2016). The root-mean-square error (RMSE) is computed when comparing time series recorded at the sampling points for the water depth (or level) and velocity. The RMSE measures the average discrepancy between the results and is calculated as:

$$\text{RMSE} = \sqrt{\frac{\sum_{i=1}^{n}[p - p_{\text{ref}}]^2}{N_s}} \tag{11}$$

where $p$ refers to the predicted flow variable (water depth, or level, or velocity) by any of the non-uniform and adaptive solvers, $p_{\text{ref}}$ indicates reference prediction by the DG2 solver on the finest uniform grid, and $N_s$ is the size of the time series. When comparing flood inundation extents, the three metrics of hit rate (H), false alarm (F) and critical success index (C) are computed (Wing et al., 2017; Hoch & Trigg, 2019). H measures the underprediction of the flood extent as the proportion of the number of inundated elements in the reference DG2 prediction that were also inundated by each solver prediction to the water depth. F measures flood extent overprediction as the proportion of the number of inundated elements of each solver that are not inundated in the reference solution. C weighs both overprediction and underprediction of the reference flood extent to assess the overall performance. All three metrics range between 0 and 1, with the best result for H and C being 1, and the best result for F being 0.

**3.1 Flood wave along a valley**

This test case has been widely used as a standard to benchmark the performance of flood inundation models and involves both rapid and gradual flood propagation stages (Neelz & Pender, 2013; Ayog et al., 2021; Shaw et al., 2021). The flood wave arises from an inflow that enters the upstream part of a 17.0 km × 0.8 km valley leading to flood propagation to the downstream over a downsloping terrain with friction effects ($n_M = 0.04$ sm$^{-1/3}$). Figure 3a shows the terrain of the valley including sampling points 1, 3 and 5 located at the upstream, middle and downstream topographic depressions. Figure 3b shows the inflow that has a high peak discharge of $Q = 3000$ m$^3$s$^{-1}$, lasting less than 2 hours. The flood wave is expected to go through two stages of rapidly and gradually propagating flows. The rapidly propagating stage occurs during the first 3.5 hours, as the water flows downhill along the valley while filling the topographic depressions to eventually reach the larger pond located downstream. The gradually propagating flow stage begins at about 5 hours to last until 30 hours, during which the flow is decelerated by friction effects to ultimately approach steady state. For configuration of the adaptive and non-uniform grid solvers, a coarsest allowed grid of 3 × 3 elements is considered with a maximum resolution level of $L = 9$ to allow the solvers to access up to the finest resolution of $R = 10$ m.

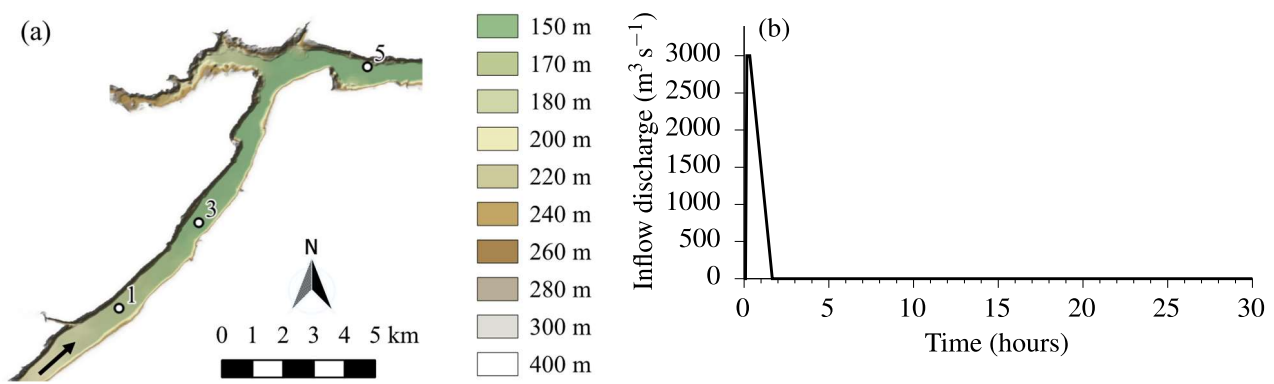

**Figure 3.** Flood wave along a valley (Section 3.1): (a) domain extent including the topography of the valley and the positions of the sampling points; (b) the inflow hydrograph imposed in the vicinity of the southern end of the domain.

### 3.1.1 Analysis of the initial non-uniform grids

The initial grids predicted by the adaptive HWFV1 and MWDG2 solvers at $\varepsilon = 10^{-3}$ are shown in Figures 4a and 4b, respectively, each including a subfigure displaying the percentage of the grid resolutions selected along the valley. HWFV1 leads to a finer grid resolution along the valley compared to MWDG2. This coarser grid prediction by MWDG2 is due to its smoother piecewise-planar representation of the topographic features, compared to the steplike piecewise-constant representation used by HWFV1 (Kesserwani & Sharifian, 2020). With HWFV1, the most frequent resolution selected along the valley is 20 m to 40 m, which is finer than the commonly used 50 m resolution for this test (Neelz & Pender, 2013; Ayog et al., 2021). In contrast, with MWDG2, the most frequent resolution selected is 80 m. It is worth noting that the predicted grids, shown in Figure 4a and Figure 4b, represent the coarsest resolution allowable during the HWFV1 and MWDG2 simulations. Figure 4c displays the non-uniform grid generated from the MRA of MWs after grading. This grid remains static-in-time when running the non-uniform solvers, and mostly involves 80 m resolution along the valley.

### 3.1.2 Analysis of velocity and water level time-series

Figure 5 shows the water level and velocity time-series recorded at the three sampling points for the adaptive and non-uniform solvers, where they are compared to the reference uniform DG2 solver predictions. A quantitative comparison is also included in Table 1 showing the respective RMSE values. As the velocity time-series (Figures 5d-5f) are recorded up to 5 hours, they are analysed first to gain insight into the solver behaviours during the rapid propagation stage.

At point 1 (Figure 5d), the earliest arrival time is recorded with the highest velocity of 2.3 ms$^{-1}$ (at around 0.5 hours when inflow forcing was still present). The MWDG2 and HWFV1 solvers and the non-uniform DG2 and FV1 solvers predict velocities that are comparable with the velocity predictions by uniform DG2. The non-uniform ACC solver, however, shows an underprediction for the velocity peak leading to the highest RMSE value. As shown in Shaw et al. (2021), before reaching point 1, the rapidly propagating flow becomes supercritical at the topographic depression upstream, which explains the underpredicted arrival time by non-uniform ACC, given its simplified local inertial formulation (De Almeida & Bates, 2013).

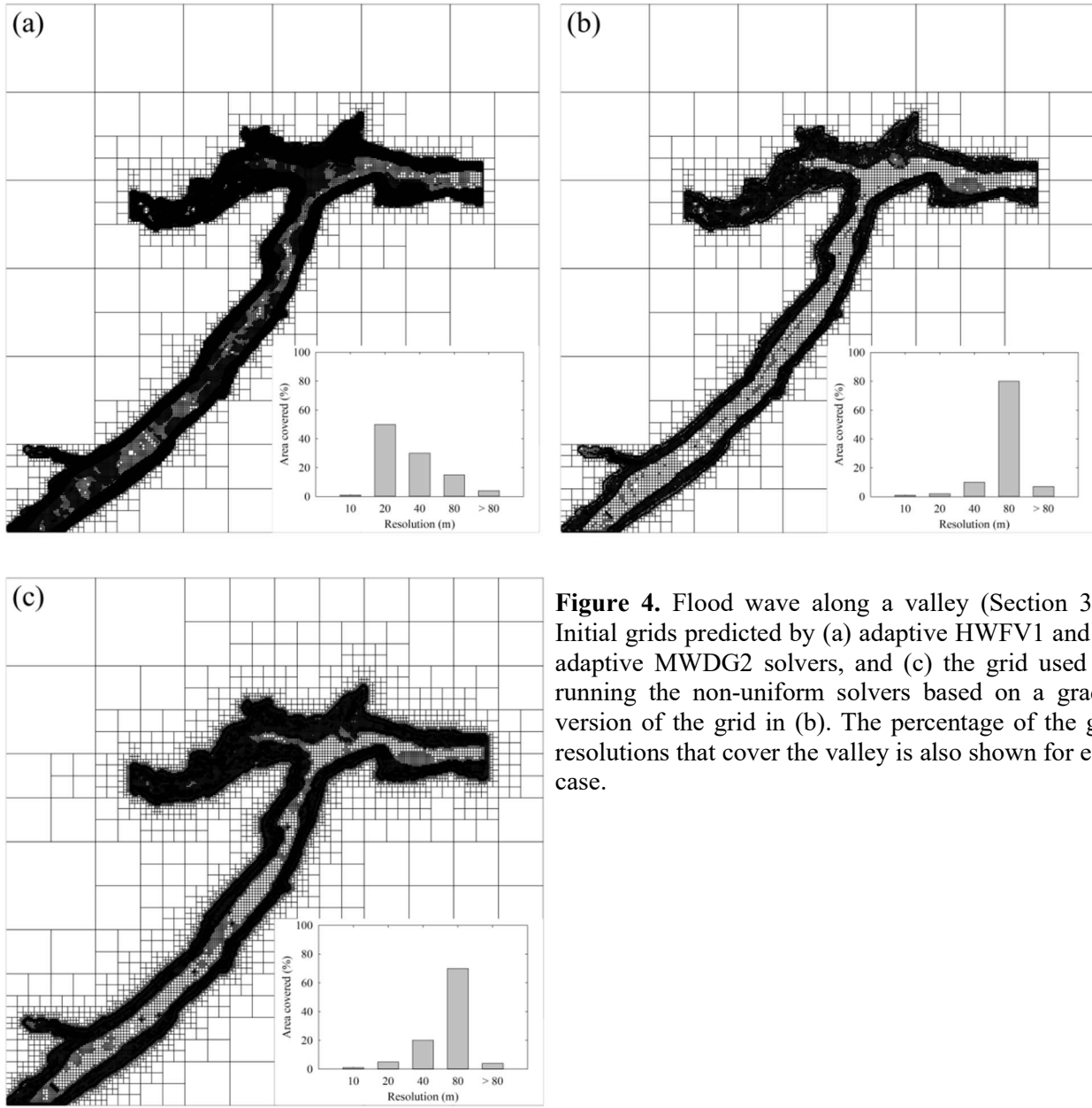

**Figure 4.** Flood wave along a valley (Section 3.1): Initial grids predicted by (a) adaptive HWFV1 and (b) adaptive MWDG2 solvers, and (c) the grid used for running the non-uniform solvers based on a graded version of the grid in (b). The percentage of the grid resolutions that cover the valley is also shown for each case.

At point 3 (Figure 5e), the wave peak, observed at point 1, has arrived without significant deceleration (by 1 hour with a peak of 2.2 ms$^{-1}$) while remaining driven by both the inflow forcing and the downsloping topography. Clearer differences between the predictions are observed: MWDG2 shows the closest agreement with uniform DG2 by capturing the small transient velocity variations, appearing after 1 hour, and by leading to an RMSE value that is one order-of-magnitude smaller than all the other RMSE values. Note that these velocity variations were also captured by an alternative MUSCL-FV2 simulation at 10 m resolution (Ayog et al., 2021). HWFV1, non-uniform FV1 and ACC overlook the presence of these velocity variations, while non-uniform DG2 could partly capture them but later lead to overly predicted velocity, leading to the highest RMSE value. The underprediction by non-uniform ACC is seen to be intensified at this point, showing around 0.7 ms$^{-1}$ difference in the predicted peak velocity.

At Point 5 (Figure 5f), as it is the farthest from the inflow, the flood is solely driven by the topography and has undergone significant deceleration by friction effects (peak velocity of 0.9 ms$^{-1}$). The MWDG2 solver shows the closest agreement with the velocities predicted by the uniform DG2 solver, while HWFV1 and non-uniform FV1 predicted slightly delayed arrival times. The velocities predicted by non-uniform ACC is close to those predicted by uniform DG2, as the local inertial equations of ACC are valid for gradually propagating flows and the locality of the ACC solver on a staggered grid delivers second-order accuracy in space (Shaw et al., 2021). The velocities predicted by non-uniform DG2 are comparable to those predicted by uniform DG2 up to 3.5 hours; however, slight noises appear in velocities after 3.5 hours that can be attributed to the approximate interpolation for the slope coefficients at the coarse element side at non-homogeneous interfaces (Section 2.2.1), leading to slight numerical disturbances. This effect is not observed with non-uniform FV1 as it merges piecewise-constant solutions across an interface separating non-homogeneous elements.

To analyse the performance of the solvers during the gradually propagation stage, the water level time series are compared in Figures 5a-5c. At points 1, the predictions by MWDG2 and HWFV1 and non-uniform DG2 and FV1 are close to reference water levels predicted by the uniform DG2, while non-uniform ACC slightly underpredicts the water levels. At point 3, all the solvers predict water levels that are close to the reference water level, though the water level predicted by non-uniform DG2 tends to be slightly lower. At point 5, MWDG2 and non-uniform FV1 and ACC perform equally well, but HWFV1 tends to overpredict the water level (by 0.2 m). This tendency may be expected with HWFV1 when it is applied to model a quantity that is steady in time, where it becomes affected by a slight accumulation of adaptation errors in line with a deeper traversal across resolution levels (see analyses in Section 3.5 in Kesserwani et al. 2019). This behaviour is not observed with MWDG2 as it deployed a much coarser grid (Figure 4), and therefore is not significantly affected by adaptation errors. The underpredictions by the non-uniform DG2 solver noted at point 3 seem to be intensified at point 5, which is likely due to the knock-on effects of the numerical disturbances arising in the velocity predictions.

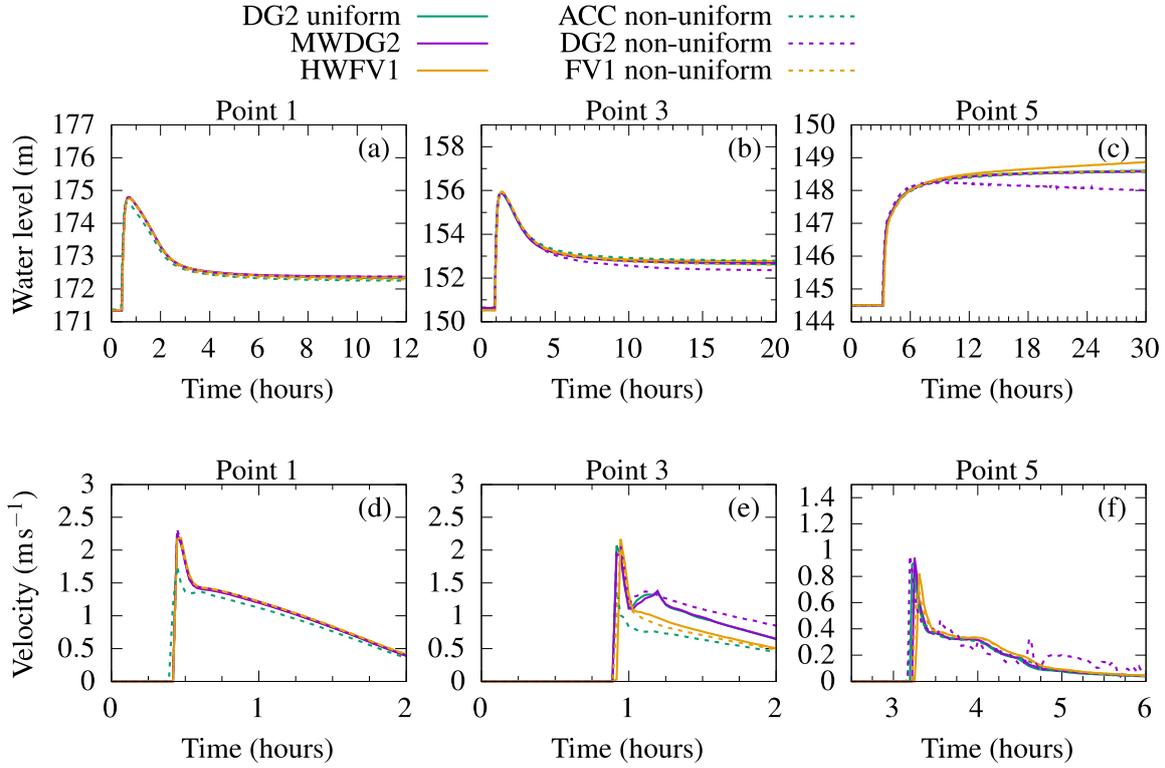

**Figure 5.** Flood wave along a valley (Section 3.1): The water level and velocity time-series predicted by the uniform DG2, adaptive MWDG2 and HWFV1 and non-uniform ACC, FV1 and DG2 solvers at sampling points 1, 3 and 5.

**Table 1.** Flood wave along a valley (Section 3.1): Root Mean Square Error (RMSE) in predicted water level (m) and velocity (ms$^{-1}$) time-series by adaptive MWDG2 and HWFV1 and non-uniform DG2, FV1 and ACC solvers, measured against the predictions by the uniform DG2 solver.

|  |  | Adaptive MWDG2 | Adaptive HWFV1 | Non-uniform ACC | Non-uniform DG2 | Non-uniform FV1 |
|---|---|---|---|---|---|---|
| Water level | Point 1 | 0.062 | 0.043 | 0.109 | 0.075 | 0.038 |
|  | Point 3 | 0.069 | 0.134 | 0.057 | 0.312 | 0.044 |
|  | Point 5 | 0.020 | 0.142 | 0.032 | 0.353 | 0.050 |
| Velocity | Point 1 | 0.016 | 0.006 | 0.106 | 0.047 | 0.006 |
|  | Point 3 | 0.008 | 0.076 | 0.079 | 0.226 | 0.063 |
|  | Point 5 | 0.027 | 0.042 | 0.013 | 0.110 | 0.032 |

### 3.1.3 Grid coarsening ability and runtime cost

Figure 6 shows the time evolution of the number of elements along the valley for the adaptive MWDG2 and HWFV1 solvers during the 30-hour long simulation, and also includes the (constant) number of elements forming the non-uniform grid (164,000 elements) and uniform grid (306,000 elements). The initial grid predicted by the MWDG2 solver has 144,000 elements while HWFV1 has 216,000 elements, around 52 % and 30% fewer than uniform DG2, respectively. During the rapidly propagating stage, the number of elements used by both MWDG2 and HWFV1 increases with roughly the same trend to reach the maximum number of the elements of 193,000 and 265,000, respectively, at 4 hours. After the flow decelerated at the start of the gradually propagating stage, the number of elements gradually decreased for both MWDG2 and HWFV1 to settle at 183,000 and 250,000, respectively, which is higher than the 162,000 elements forming the non-uniform grid. This is expected with HWFV1 as its initial grid already has a higher number of elements. The grid predicted by MWDG2, though initially has less than 162,000 elements, does not settle to a lower number, being subjected to refinement in response to the flow dynamics.

The CPU runtime costs required by the solvers to accomplish the 30-hour long simulation are shown in Table 2. Uniform DG2 demands the highest runtime cost of 144 hours. MWDG2 and HWFV1 are 2.3 and 4.5 times faster to run, costing 61 and 32 hours, respectively. Still, they remain more expensive to run than all the non-uniform solvers that are not subjected to dynamic adaptation overhead costs. For the non-uniform solvers, as expected, the DG2 is the costliest, taking 22 hours, while the costs for running FV1 and ACC are 5.6 and 5.1 hours, respectively. Hence, non-uniform DG2, FV1 and ACC offer speed-ups equal to 6.5, 25.7 and 28.2 over uniform DG2, respectively.

Overall, this test case further confirms the previous findings in Kesserwani and Sharifian (2020), that when modelling a rapidly propagating flood, MWDG2 and HWFV1 are cheaper alternatives to retrieve the modelling quality of the expensive uniform DG2 solver. MWDG2 might be particularly preferred when the aim of the modelling is to more accurately capture velocity transients. On the other hand, when modelling a gradually propagating flood, the overhead cost of the adaptation process in MWDG2 and HWFV1 may be unnecessary as it increases the total runtime cost of simulation to deliver outcomes that are retrievable by the non-uniform solvers at a much cheaper runtime cost. Amongst the non-uniform solvers, FV1 seems to deliver the most comparable quality to uniform DG2 for this test with rapidly propagating flows, with a considerably fewer number of elements and at a competitive runtime cost to the less complex ACC solver. The non-uniform DG2 solver's velocity predictions can be affected by slight numerical disturbances, but their effects remain insignificant on flood extent and water level predictions.

As the configuration of this test case only involved a single stream channel with an (almost) unidirectional flow, further scenarios featured with more realistic 2D floodplain flows are explored next, to evaluate the solvers in modelling gradually propagating flood inundation.

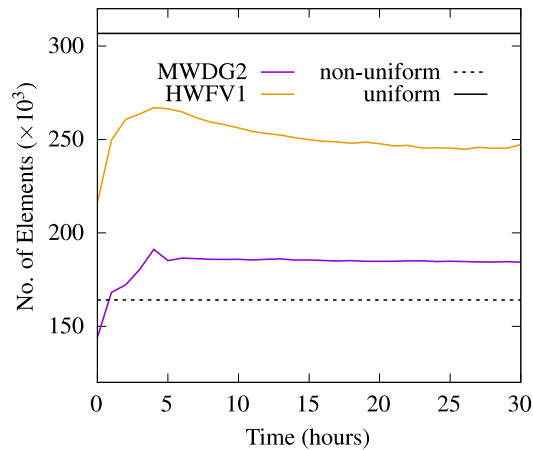

**Figure 6.** Flood wave along a valley (Section 3.1): Variation of the number of elements along the valley used by the MWDG2 and HFV1 solvers over the 30-h long simulation. The numbers of elements used in the uniform and the MW-based non-uniform grids are marked as the solid and dotted black horizontal lines, respectively.

**Table 2.** Flood wave along a valley (Section 3.1): Solver runtimes for uniform DG2, adaptive MWDG2 and HWFV1, and non-uniform DG2, FV1 and ACC solvers, over the 30-hour long simulation.

| Solver | Uniform DG2 | Adaptive MWDG2 | Adaptive HWFV1 | Non-uniform DG2 | Non-uniform FV1 | Non-uniform ACC |
|---|---|---|---|---|---|---|
| Runtime | 144 hours | 61 hours | 32 hours | 22 hours | 5.6 hours | 5.1 hours |

**3.2 Hypothetical flood propagation and inundation in Thamesmead**

The performance of the adaptive and non-uniform solvers is assessed when applied to reproduce the Thamesmead hypothetical flood inundation scenario (Liang et al., 2008; Wang and Liang, 2011; Vacondio et al., 2012). The test case involves a gradually propagating flood over a complex terrain in the Thamesmead district. The domain is located in a low-lying and densely populated area downstream of London's main flood defence, the Thames Barrier. The flood is assumed to occur from a 150 m long breach in the defence structure, through which the flood inflows to a 9 km × 4 km floodplain with a uniform Manning parameter of $n_M = 0.035$ sm$^{-1/3}$. The aerial view of the floodplain is shown in the left panel of Figure 7, which also includes the location of the inflow breach and four sampling points where water level and velocity time series are recorded. The inflow is represented by the discharge hydrograph shown in the right panel of Figure 7 that lasts for 10 hours with a peak of $Q = 200$ m$^3$s$^{-1}$ occurring between the second and fourth hours of flooding. The flood wave is expected to pass a narrow channel that cuts through a railway embankment located across most of the domain from southwest to northeast. The topography of the floodplain is represented by a 10 m resolution digital terrain model (DTM) dataset (Hou et al., 2018), which includes complex features like waterways, channels and the railway embankment. Capturing these features demands a much finer non-uniform grid than the previous test case (Section 3.1.1). The adaptive MWDG2 and HWFV1 solvers are run on a coarsest allowed grid of 2 × 1 elements and a maximum resolution level $L = 9$, to allow the solvers access up to the finest resolution of $R = 10$ m, while the non-uniform DG2, FV1 and ACC solvers are run on the graded non-uniform grid predicted by the MRA of MWs.

**3.2.1 Analysis of the initial non-uniform grids**

Figures 8a and 8b show the initial grids predicted by HWFV1 and MWDG2, driven by their piecewise-constant and piecewise-planar representation of the DTM, respectively. HWFV1 predicts a more refined grid with 206,000 elements, which is only 3% fewer than the total number of elements on the uniform grid counterpart, with 211,000 elements (excluding the areas outside of the floodplain). MWDG2 only selects the finest resolution in the vicinity of the sharpest features, while using coarser resolutions in the smoother areas due to its piecewise-planar topography representation. Hence, the initial grid predicted by MWDG2 used 150,000 elements, 30% fewer elements than the uniform grid. The graded non-uniform grid generated by the MRA of MWs is shown in Figure 8c and is composed of 192,000 elements, which is 10% fewer than the total elements on the uniform grid. Since the embankment channel area is fully refined in all three grids (Figures 8a to 8c), the choice of $\varepsilon = 10^{-3}$ is deemed appropriate. This is particularly important as the channel area is the only route to convey the floodwater into the southern side of the floodplain (Hou et al., 2018).

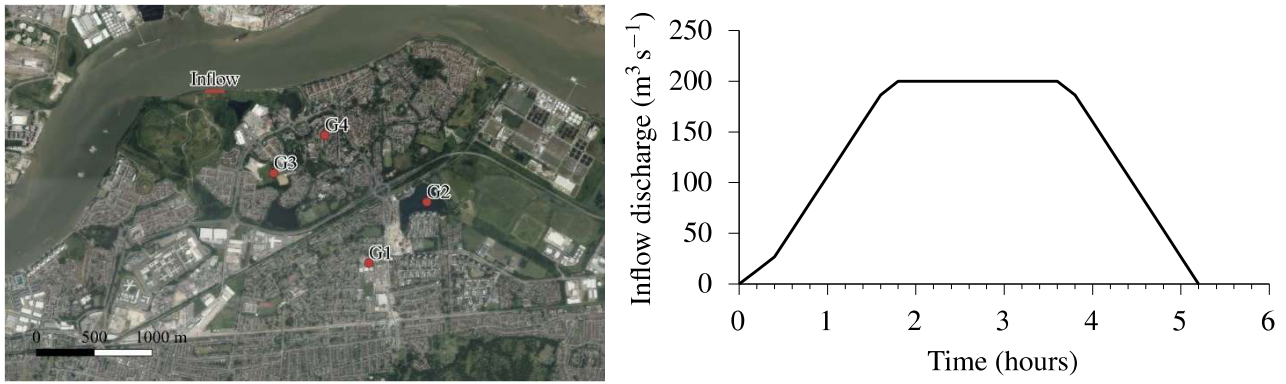

**Figure 7.** Hypothetical flood propagation and inundation in Thamesmead (Section 3.2): (left) 9 km × 4 km extent and the positions of the sampling points and the hypothetical breach in the flood defence structure; (right) the inflow hydrograph. Map data ©2020 Google.

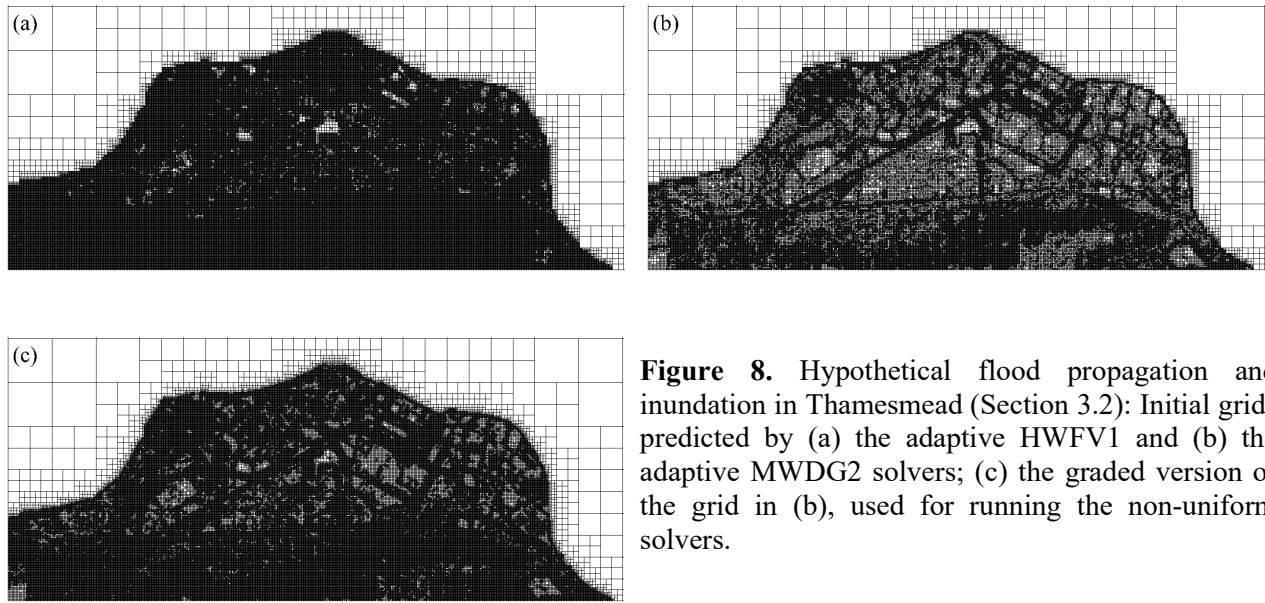

**Figure 8.** Hypothetical flood propagation and inundation in Thamesmead (Section 3.2): Initial grids predicted by (a) the adaptive HWFV1 and (b) the adaptive MWDG2 solvers; (c) the graded version of the grid in (b), used for running the non-uniform solvers.

### 3.2.2 Analysis of velocity and water level time-series

Figure 9 compares the water level and velocity time-series recorded at the sampling points G1 to G4 for the adaptive and non-uniform solvers, to the reference time series predicted by the uniform DG2 solver, while their respective RMSE values are listed in Table 3. The predicted time-series are discussed in a particular order for the sampling points, starting from those located closest to the inflow to the farthest ones, first for the water levels and then for the velocity.

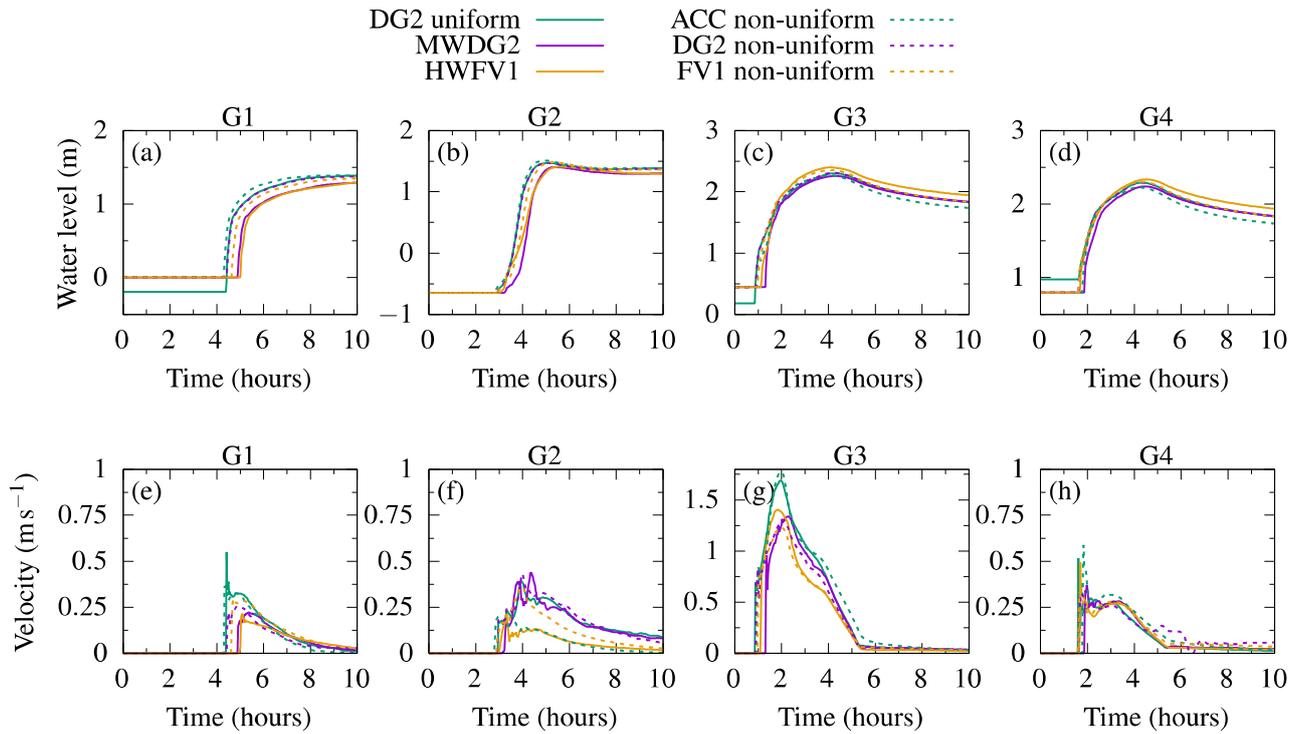

**Figure 9.** Hypothetical flood propagation and inundation in Thamesmead (Section 3.2): The water level and velocity time-series predicted by the uniform DG2, adaptive MWDG2 and HWFV1 and non-uniform ACC, FV1 and DG2 solvers at sampling points G1, G2, G3 and G4.

At points G3 (Figure 9c) and G4 (Figure 9d), the water levels predicted by non-uniform DG2 and FV1 are the closest to reference water levels predicted by uniform DG2, yielding the lowest RMSE values (around 0.04). Until 4 hours, non-uniform ACC closely trails the water level predicted by uniform DG2; but later, it shows a slight underprediction by a maximum of 0.1 m at G4 where it underperforms non-uniform FV1 and DG2 that have lower RMSE values. MWDG2 predicts a delay in the arrival time of around 0.4 and 0.2 hours at G3 and G4, respectively, leading to the highest RMSE value (around 0.09). This underperformance for MWDG2 may be expected given its much coarser grid compared to the other solvers (recall Section 3.2.1). Nonetheless, MWDG2 remains able to trail the reference water level of DG2 uniform after 5 hours. HWFV1, despite its finer grid, slightly overpredicts the water level by 0.1 m, after 2 hours until the end of the simulation, and leads to an RMSE value that is higher than non-uniform FV1 at both points G3 and G4. This overprediction might be due to the accumulation of the adaptation errors as the HWFV1 solver entails a deeper traversal across the resolution levels due to its finer resolution (Section 3.1.2)

Point G2 is located inside a lake at the south of the railway embankment. Decelerated by the friction effects at the north of the embankment, the water reaches the lake at 3.5 hours and accumulates there until the lake becomes full by 5 hours. The lake, being full, is then overflowed and the water level within remains almost constant. As shown in Figure 9b, the water levels predicted by non-uniform DG2, FV1 and ACC are the closest to water levels predicted by uniform DG2, leading to RMSE values of 0.004, 0.06, and 0.019, respectively. The RMSE values for MWDG2 and HFV1 are relatively bigger, i.e., 0.147 and 0.101, respectively. This indicates that the adaptive solvers underperform the non-uniform solvers for this test: MWDG2 predicts a delay in the wave arrival time and HWFV1 significantly underpredicts the water level between 4 to 5 hours. However, MWDG2 and HWFV1 predictions get closer to the uniform DG2 water level after the overflow from the lake.

At Point G1, MWDG2 and HWFV1 lead to even bigger deviations in water level predictions (Figure 9a), showing the least agreement with the uniform DG2: a 0.6-hour delay in the arrival time and around 0.1 m underprediction throughout, with RMSE values of 0.139 and 0.160, respectively. Non-uniform DG2, FV1 and ACC have relatively smaller RMSE values of 0.075, 0.098 and 0.089, respectively, all predicting closer water levels to the reference uniform DG2 prediction, with FV1 having the biggest RMSE value due to its slightly delayed arrival time.

For the velocity predictions, the discrepancy between the solver predictions is more prominent, although not as conclusive as the previous test case, as this test only involves a gradually propagating flow with a relatively smaller velocity magnitude. At point G3 (Figure 9g), non-uniform ACC delivers the closest velocity peak to that of uniform DG2 and leads to the lowest RMSE value of 0.049; though it overpredicts the velocity after 4 hours. Comparatively, the other solvers underpredict the velocity peak with higher RMSE values, around 0.1, except for non-uniform DG2 that leads to the second-best predictions after ACC with an RMSE value of 0.078. MWDG2 predictions have the highest RMSE value of 0.128, which may be due to its delayed wave arrival time on a much coarser grid (recall Section 3.2.4). HFV1 and non-uniform FV1, despite their finer grids, underpredict the velocity during 2 to 4 hours. At point G4, however, non-uniform ACC leads to the highest RMSE value of 0.034, while HFV1 and non-uniform FV1 deliver the lowest ones, around 0.022. At this point only MWDG2 and non-uniform DG2 could capture the small velocity variations predicted by uniform DG2 around 2.5 hours, suggesting the ability of the DG2-based solvers in predicting small velocity variations as observed earlier in Section 3.1.2. Their slightly higher RMSE values, around 0.027, might be expected as, at point G4, MWDG2 also predicts a delayed arrival time, by 0.3 hours, and non-uniform DG2 slightly overpredicts velocity after 5 hours. Similar is observed at point G2: MWDG2 and non-uniform DG2 are the only solvers that capture the small velocity transient predicted by uniform DG2, leading to the lowest RMSE values of 0.025 and 0.011, respectively, despite MWDG2 predicting 0.5 hours delay in the wave arrival time. HFV1 and non-uniform ACC and FV1 underpredict the velocity leading to RMSE values around 0.6 for ACC and HFV1 and around 0.3 for non-uniform FV1. At point G1, MWDG2 and HFV1 predict the most deviated velocities, given their lagged arrival times, with RMSE values around 0.05; whereas the non-uniform solvers predict much closer velocities to those predicted by uniform DG2, with RMSE values around 0.025. These analyses show that the non-uniform solvers can perform better than the adaptive solvers to more accurately capture wave arrival times at the gauge points, in particular those located far from the inflow. The non-uniform DG2 solver leads to the lowest RMSE values overall and could capture small velocity variations as opposed to non-uniform ACC and FV1.

**Table 3.** Hypothetical flood propagation and inundation in Thamesmead (Section 3.2): Root Mean Square Error (RMSE) in predicted water level (m) and velocity (ms$^{-1}$) time-series by adaptive MWDG2 and HWFV1 and non-uniform DG2, FV1 and ACC solvers, measured against the predictions by the uniform DG2 solver.

|  |  | Adaptive MWDG2 | Adaptive HWFV1 | Non-uniform ACC | Non-uniform DG2 | Non-uniform FV1 |
|---|---|---|---|---|---|---|
| Water level | G1 | 0.139 | 0.160 | 0.089 | 0.075 | 0.098 |
|  | G2 | 0.147 | 0.101 | 0.019 | 0.004 | 0.060 |
|  | G3 | 0.096 | 0.094 | 0.068 | 0.047 | 0.060 |
|  | G4 | 0.069 | 0.060 | 0.066 | 0.041 | 0.042 |
| Velocity | G1 | 0.047 | 0.052 | 0.023 | 0.027 | 0.033 |
|  | G2 | 0.025 | 0.063 | 0.064 | 0.011 | 0.033 |
|  | G3 | 0.128 | 0.099 | 0.049 | 0.078 | 0.109 |
|  | G4 | 0.027 | 0.023 | 0.034 | 0.028 | 0.022 |

### 3.2.3 Analysis of the flood inundation maps

Figure 10 shows the flood inundation maps predicted by the uniform DG2 and the adaptive and non-uniform solvers after 10 hours. A quantitative comparison of flood inundation extents is provided in Table 4 based on the hit rate (H), false alarm (F) and critical success index (C) metrics, using the uniform DG2 prediction as a reference. The water depth maps in Figure 10 show that the adaptive and non-uniform solver predict inundation depths that are comparable with the water depths predicted by the uniform DG2 solver. Clearer differences are observed among the solver predictions for the flood inundation extent. MWDG2 predicts a relatively narrower extent to that of uniform DG2, e.g., the southeast and southwest corners of the flooded area. These narrower predictions are expected with MWDG2 as it led to delayed arrival times (recall Figure 9) given its aggressive coarsening in its grid resolution (see also Section 3.2.4). These underpredictions are also reflected in the hit rate of H = 0.86 which is the lowest among the solvers. The MWDG2 also leads to a relatively low false alarm of F = 0.005 and a critical success index of C = 0.856. Compared to MWDG2, the HWFV1 shows relatively less tendency to underpredict the flood extent by having a higher hit rate of H = 0.89. However, HWFV1 still shows a narrower flood extent than the reference uniform DG2, in the shallow channels located at the east of the floodplain, which might be due to less accurate representation of the channel topography by its piecewise-constant accuracy. The higher false alarm of F = 0.084 with HWFV1, compared to MWDG2, indicates its tendency to overpredict the flood extent. HWFV1 also leads to a comparatively lower critical success index of C = 0.822. The non-uniform DG2, FV1 and ACC solvers predict better extents than the adaptive solvers that are much closer to the extent predicted by the uniform DG2 solver, with only negligible differences in the southernmost region of the flooded area: non-uniform DG2 leads to the highest critical success index of C = 0.912, followed by non-uniform FV1 and ACC with C = 0.891 and 0.883, respectively. The better performance for the non-uniform solvers, compared to adaptive ones, may be expected as their grid was initially generated using the MRA of MWs applied to smooth piecewise-planar representation of DTM data, and then subjected to grading that leads to increased finer resolutions (Figure 8). On the latter grid,

DG2 is expected to perform slightly better for smooth flows as it stores and evolves piecewise-planar solutions whereas FV1 and ACC use piecewise-constant solutions.

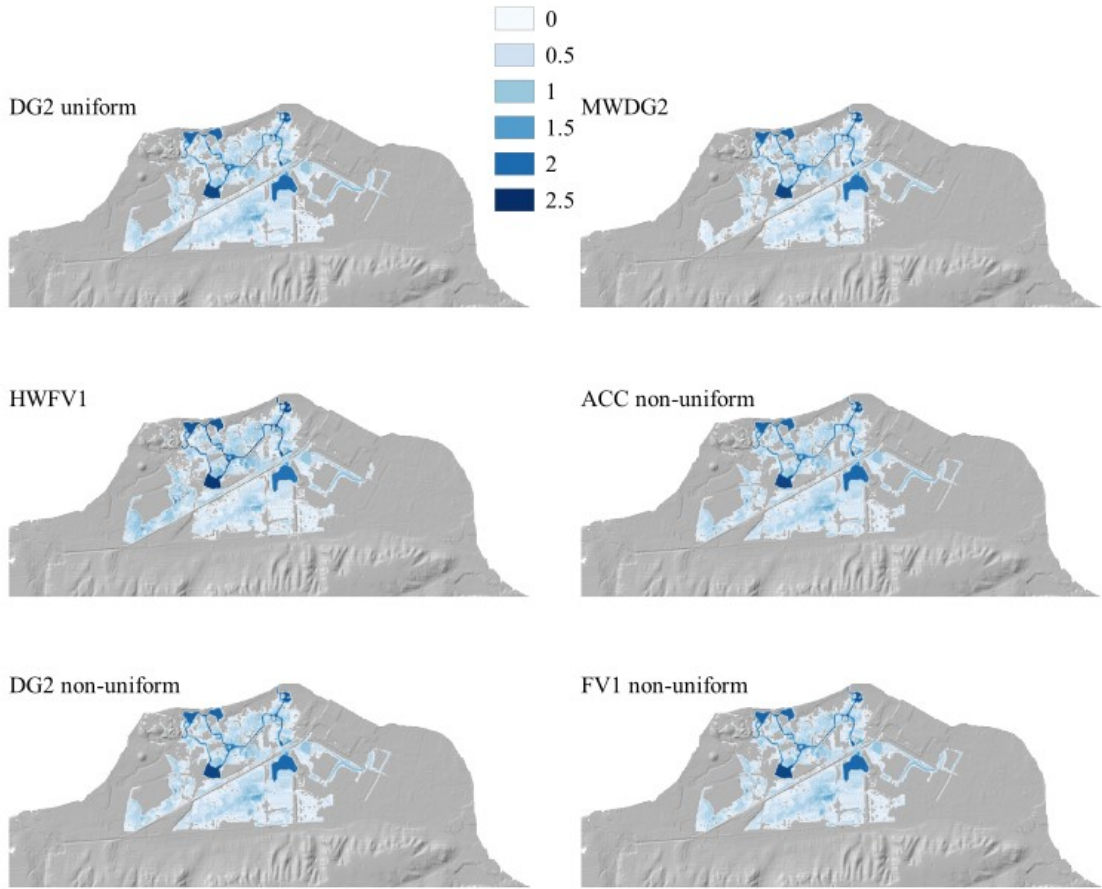

**Figure 10.** Hypothetical flood propagation and inundation in Thamesmead (Section 3.2): Final predicted flood inundation maps.

**Table 4.** Hypothetical flood propagation and inundation in Thamesmead (Section 3.2): Hit rate (H), false alarm ratio (F) and critical success index (C) for the flood maps predicted by the adaptive MWDG2 and HWFV1 and non-uniform DG2, FV1 and ACC solvers at 10 h, against the reference predictions by the uniform DG2 solver.

| Solver | Adaptive MWDG2 | Adaptive HWFV1 | Non-uniform DG2 | Non-uniform FV1 | Non-uniform ACC |
|---|---|---|---|---|---|
| H | 0.860 | 0.890 | 0.941 | 0.928 | 0.903 |
| F | 0.005 | 0.084 | 0.032 | 0.028 | 0.044 |
| C | 0.856 | 0.822 | 0.912 | 0.891 | 0.883 |

### 3.2.4 Grid coarsening ability and runtime cost

Figure 11 shows the variations of the number of elements used by the adaptive MWDG2 and HWFV1 solvers during the 10-hour long simulations, along with the (constant) number of elements forming the non-uniform and uniform grids. The number of elements used by the HWFV1 solver is initially 3% fewer than the uniform and 7% more than the non-uniform grid counterparts. With the evolution of the flooding their number of elements increases, so that after 5 hours HWFV1 only has 1% fewer elements than the uniform and 8% more than the non-uniform grid. The grid of MWDG2 initially uses 30% fewer elements than the uniform and 21% fewer elements than the non-uniform grids. Its number of elements increases to reach its maximum after 6 hours, which is still 25% fewer than the number of elements on the uniform grid and 17% fewer than the non-uniform grid, suggesting that it is much more efficient for dynamic grid adaptation without a tendency to overly refine. Compared to the previous test case (Section 3.1.3), HWFV1 and MWDG2 lead to less reductions in the number of elements, leading to lower speed-up ratios. Table 4 lists the CPU runtimes consumed by the solvers to complete the 10-hour long simulations. As expected, the uniform DG2 solver results in the highest runtime cost of 15.1 hours, while MWDG2 and HWFV1 require 9.6 and 7.4 hours, respectively, making them 1.6 and 2 times faster than uniform DG2. The runtime costs of non-uniform DG2, FV1 and ACC solvers are 8.7, 1.8 and 1.35 hours, which are 1.7, 8.4 and 11.2 times faster than the uniform DG2 solver, respectively.

In summary, the results for this test case further suggest that the non-uniform solvers are better alternatives to the adaptive solvers to simulate gradually propagating flooding flows, allowing to avoid the overhead cost associated with dynamic (multi)wavelet-based grid adaptation. To model such flooding scenarios, non-uniform DG2 is here found to be a better alternative to capture small-scale velocity transient, but otherwise FV1 and ACC are more suited choices to lessen further the runtime cost. Compared to the previous test case (Section 3.1), the ACC solver leads to a higher quality of predictions as here the flow remains in the lower range of subcritical flows ($Fr < 0.5$), which is within the validity scope of its local inertial equations (De Almeida and Bates, 2013).

Even though this test case explored the performance of the adaptive and non-uniform solvers in modelling gradually propagating floodplain flows, it still lacked some aspects of real-world flooding scenarios: only includes a hypothetical defence breach, one inflow source, and a DTM excluding urban features like buildings. This smoothness in the DTM resulted in aggressive coarsening of the grid by the MWDG2 solver, causing a delay in the propagation of the flood flow. All these missing aspects are explored in the next test case.

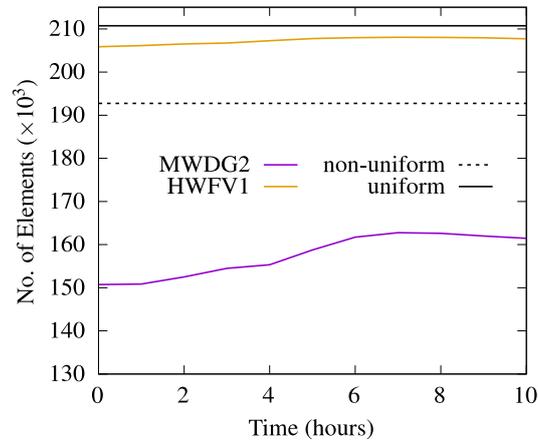

**Figure 11.** Hypothetical flood propagation and inundation in Thamesmead (Section 3.2): Variation of the number of elements used by the MWDG2 and HFV1 solvers over the 10 h long simulation. The numbers of elements used in the uniform and the MW-based non-uniform grids are marked as the solid and dotted black horizontal lines, respectively.

**Table 5.** Hypothetical flood propagation and inundation in Thamesmead (Section 3.2): Solver runtimes for Uniform DG2, Adaptive MWDG2 and HWFV1, and non-uniform DG2, FV1 and ACC solvers, over the 10 h long simulation.

| Solver | Uniform DG2 | Adaptive MWDG2 | Adaptive HWFV1 | Non-uniform DG2 | Non-uniform FV1 | Non-uniform ACC |
|---|---|---|---|---|---|---|
| Runtime | 15.1 hours | 9.6 hours | 7.4 hours | 8.7 hours | 1.8 hours | 1.35 hours |

**3.3 Carlisle 2005 flooding**

The city of Carlisle is located in the downstream Eden Catchment in Northwest England, at the confluence of River Eden and its two tributaries, Rivers Petteril and Caldew. In early January 2005, heavy rainfall over the Cumbrian mountains led to high water levels in these rivers causing widespread inundation throughout the city due to the overflow of a number of flood defences. Approximately 1,934 properties were directly flooded, and three people died (Fewtrell et al., 2011). This flooding scenario is of particular interest as it happens over an entire city with realistic urban topography from multiple fluvial sources of inflow. Moreover, the availability of post-event water and wrack marks (Neal et al., 2009) makes it a suitable case study for assessing the performance of flood models (Horritt et al., 2010; Fewtrell et al., 2011; Liu and Pender, 2013; Kabir et al., 2020). Figure 12 shows an aerial view of the study domain which covers about 14.5 km² of Carlisle's area, along with the location of the 15 sampling points where water depth time series are recorded. The three-day long flooding is initiated from three inflows at the upstream points of Rivers Eden, Petteril and Caldew (marked in Figure 12), using the hydrographs depicted in Figure 13. The topography of the floodplain is represented by a 5 m resolution DEM, provided by the Environment Agency of England and Wales. This resolution was shown to be sufficient to represent the urban features required (Xing et al., 2019; Xia et al., 2019). A uniform Manning parameter of $n_M = 0.055$ sm$^{-1/3}$ is considered for the floodplain. The adaptive solvers are run on a coarsest allowed grid of $4 \times 3$ elements with a maximum level $L = 8$, to allow these solvers to access up to the finest resolution of $R = 5$ m and the non-uniform solvers are run on the graded grid generated by the MRA of MWs.

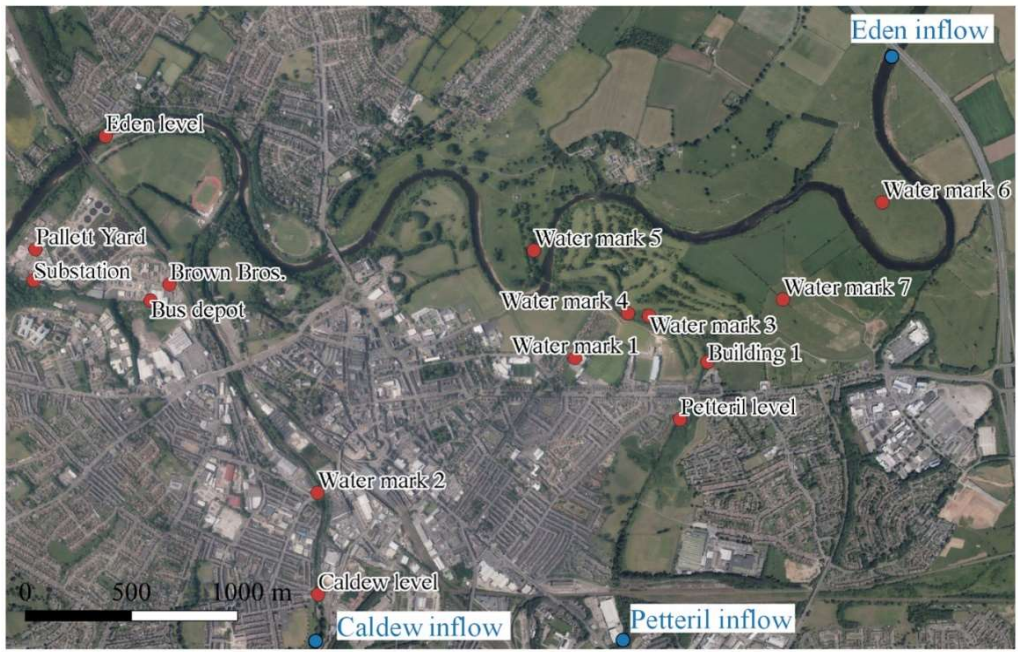

**Figure 12.** Carlisle 2005 Flooding (Section 3.3): the 14.5 km$^2$ domain extent and the positions of the sampling points and the inflows at the upstreams of Rivers Eden, Petteril and Caldew. Map data ©2020 Google.

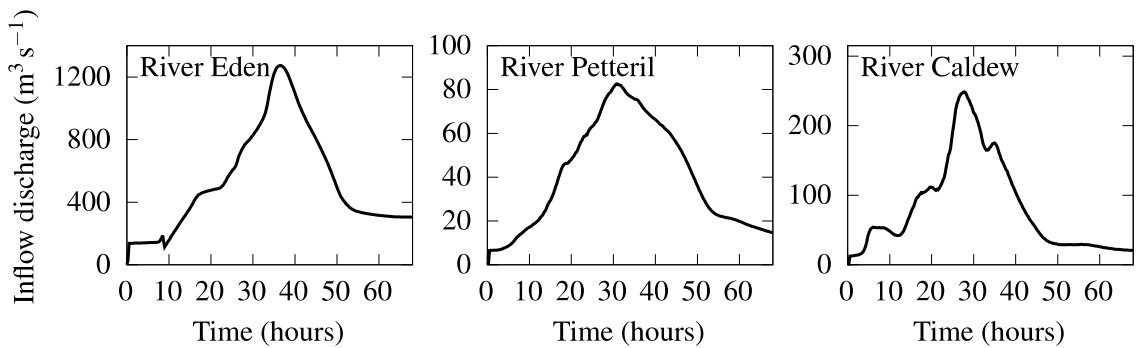

**Figure 13.** Carlisle 2005 Flooding (Section 3.3): The inflow hydrographs imposed at the upstreams of Rivers Eden, Petteril and Caldew.

### 3.3.1 Analysis of the initial non-uniform grids

Figures 14a and 14b show the initial grids predicted by the HWFV1 and MWDG2 solvers, respectively. The HWFV1 solver is shown to generate an overly refined grid with 560,000 elements, which is only 4% fewer than the total number of 581,000 elements on the uniform grid. The MWDG2 solver, however, delivers a more sensible grid resolution adaptation: more aggressive coarsening, up to a resolution of 20 m, is observed over the rural areas that cover most of the northern side of the floodplain; whereas, the topographic features like Eden river's channel banks, the motorway along the eastern edge of the domain and the rural roads are more selectively refined. The grid has 435,000 elements, which is 25% fewer elements than the uniform grid. The graded non-uniform generated by the MRA of MWs is shown in Figure 14c. Being made of 506,000 elements, it uses 13% fewer elements than the uniform grid. Note that the reduction in the number of elements predicted by the HWFV1 and MWDG2 solvers and on the latter non-uniform grid is roughly in the same order as the counterparts from the previous test case (recall Section 3.2.1).

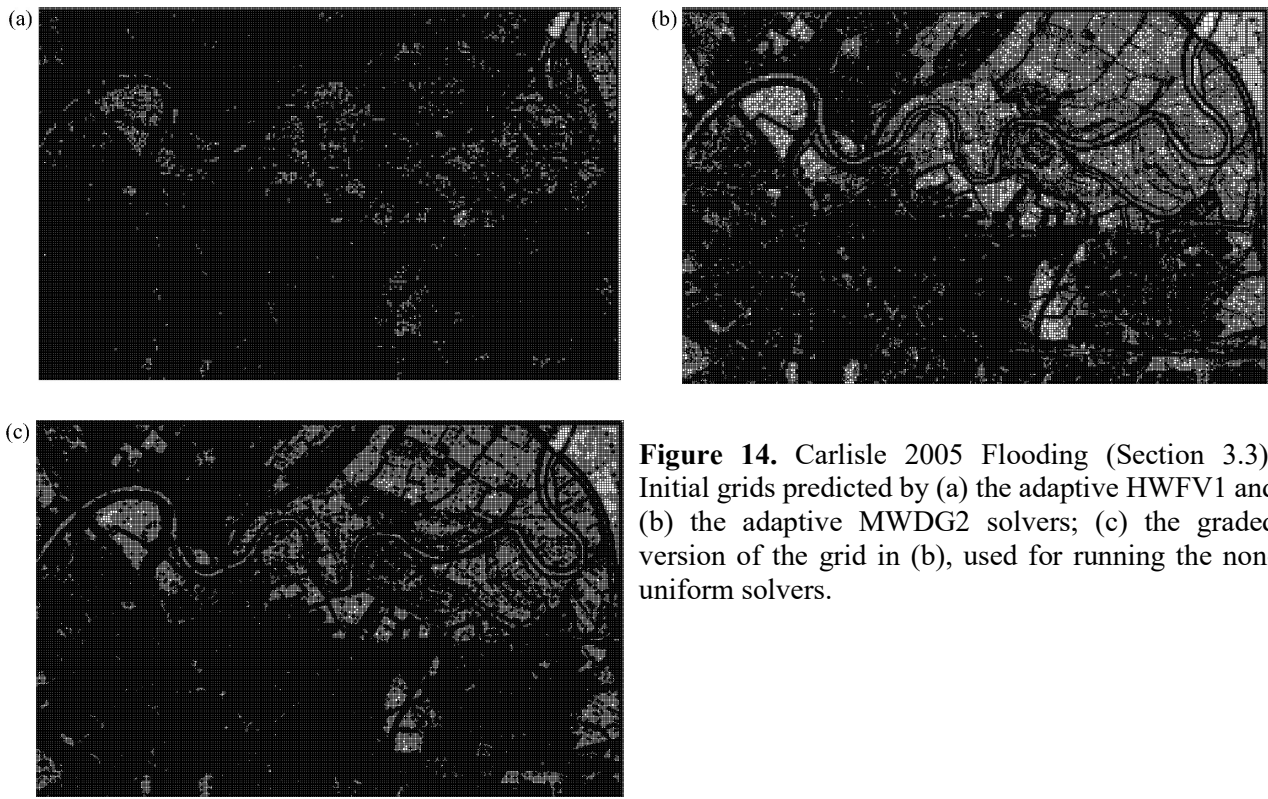

**Figure 14.** Carlisle 2005 Flooding (Section 3.3): Initial grids predicted by (a) the adaptive HWFV1 and (b) the adaptive MWDG2 solvers; (c) the graded version of the grid in (b), used for running the non-uniform solvers.

**3.3.2 Analysis of velocity and water level time-series**

Figure 15 compares the water level time-series recorded at 15 sampling points for the adaptive and non-uniform solvers, to the reference time-series predicted by uniform DG2. The samplings points are distributed over the whole domain, being located at the river channels (Sheepmount, Botcherby Bridge and Denton Holme), urban areas (Pallet yard, Substation, Brown Bros, Bus depot, Water mark 1 and Water mark 2) and rural areas (Building 1, and Watermarks 3 to 7). The three sampling points of Sheepmount, Botcherby Bridge and Denton Holme are located at the flow measurement gauges of the rivers Eden, Petteril and Caldew, respectively (see Figure 12), and therefore include observation data (Neal et al., 2009) shown in Figure 15.

As seen in the time series, the MWDG2 generally leads to delayed arrival times depending on the positions of the sampling points. Due to these delays, the water depths predicted by MWDG2 remain lower than the predictions by the uniform DG2 up to the peak of the flood. The delays, also seen in the previous test case (Section 3.2.2), are expected with MWDG2 as its grid is coarser compared to the other solvers. Nevertheless, once the flood begins to recede, the MWDG2 solver is able to trail the reference water depth predictions, likely due to the capability of the DG2-based solvers to better capture vanishing velocities (Ayog et al., 2021). The non-uniform DG2 solver performs slightly better than the MWDG2 solver by trailing the predictions of the uniform DG2 solver with a closer agreement. The superior performance of the non-uniform DG2 solver is likely due to its finer resolution grid, as also observed in the previous test case (Section 3.2.2). The HWFV1 and non-uniform FV1 solvers are able to model the trends of rising and falling limbs but lead to shorter arrival times and tend to overpredict the water depth. These overpredictions, which depending on the positions of the points, vary between 0.2 to 0.9 m, are in line with previous study findings using uniform FV1 solvers (Liu and Pender, 2013; Shaw et al., 2021). The non-uniform ACC generally leads to a better agreement with the uniform DG2, compared to the HWFV1 and non-uniform FV1, especially at the sampling points located in rural areas. The same behaviour has been previously reported in Shaw et al. (2021), where the uniform ACC solver predicted more accurate hydrographs than uniform FV1 for a similar real-world fluvial flood simulation.

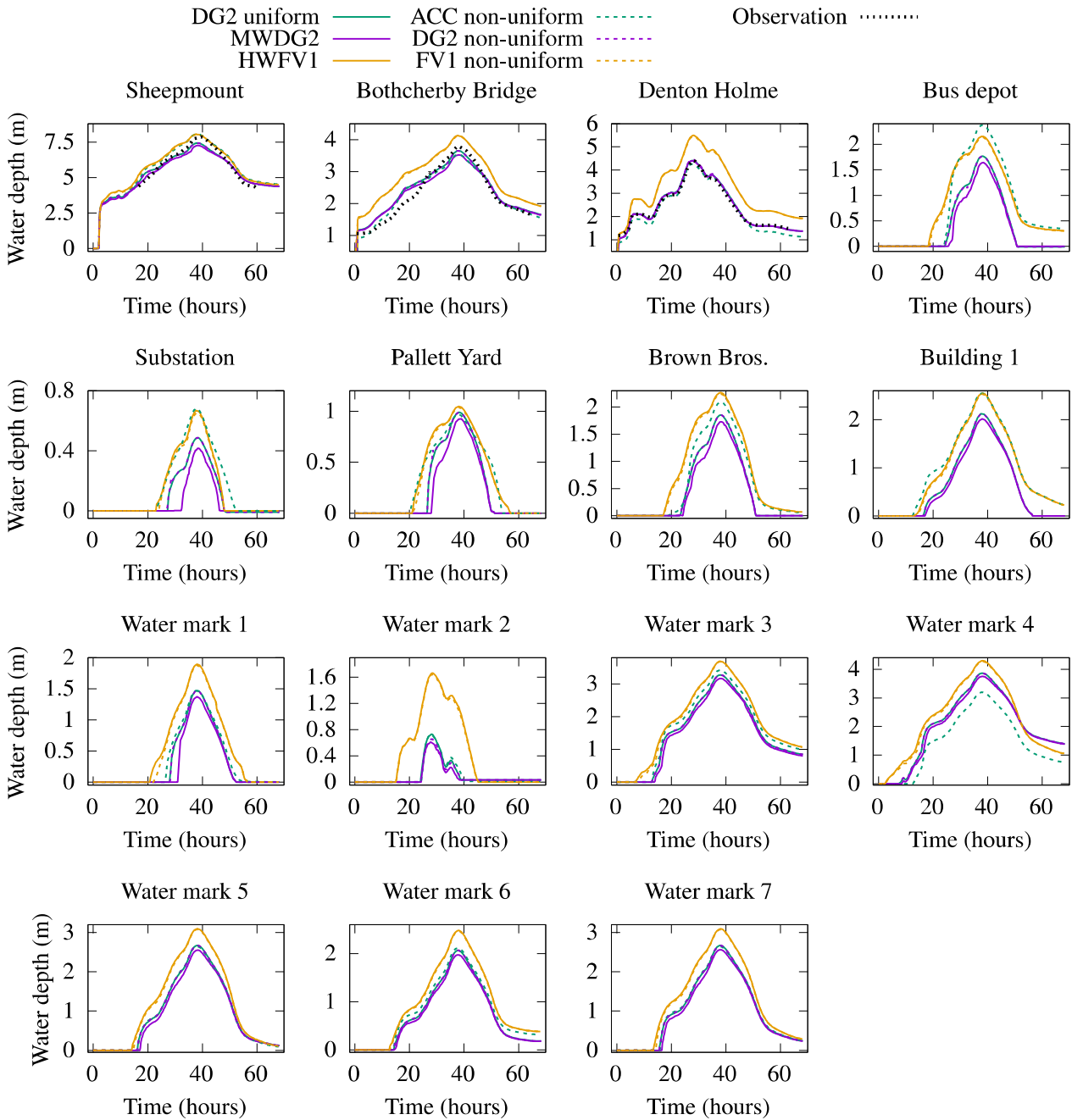

**Figure 15.** Carlisle 2005 Flooding (Section 3.3): The water level and velocity time-series predicted by the uniform DG2 solver, adaptive MWDG2 and HWFV1 solvers and ACC, FV1 and DG2 solvers on the MW-based non-uniform grid at 15 sampling points marked in Figure 12.

### 3.3.3 Analysis of the flood inundation maps

The trends of the flooding and recession in the hydrographs (Figure 15) suggests that the peak volume over the floodplain happened by 40 hours, leading to the widest inundation extent. Comparisons between the predicted flood inundation maps are therefore provided at this output time, in Figure 16, for prediction made by the adaptive and non-uniform solvers and the reference uniform DG2 solver. Their respective hit rate (H), false alarm (F) and critical success index (C) metrics are provided in Table 6.

MWDG2 and non-uniform DG2 predict inundation extents that are the closest to reference extent, subject to slight underpredictions over the urban banks of rivers Caldew and Petteril. These underpredictions are not major as the hit rates of the MWDG2 and non-uniform DG2 are 0.97 and 0.98, respectively. The false alarms of 0.005 and 0.008 also indicate that the two solvers only slightly overpredict the flood extent. The combination of the high hit rates and the low false alarms leads to the best overall performance among the solvers for the MWDG2 and non-uniform DG2 with the critical success index of 0.97. With HWFV1 and non-uniform FV1, the deviations from the reference flood extent are more pronounced, leading to a wider flood extent, especially over the urban area in the west of river Caldew. These overpredictions of the flood extent are expected as the FV1-based solvers resulted in higher water depths compared to the other solvers (Figure 15). Even though the hit rates resulting from HWFV1 and non-uniform FV1 are both close to unity (i.e., 0.99), their higher false alarms (0.10 for HWFV1 and 0.08 for non-uniform FV1) suggest that they lead to the worst performances overall, as also indicated by their lower critical success indexes (C = 0.89 for HWFV1 and C = 0.90 for non-uniform FV1). Comparatively, the non-uniform ACC solver leads to a better agreement with the reference DG2 prediction than the HWFV1 and non-uniform FV1, as suggests its higher critical success index of C = 0.92. Although, it tends to predict a wider flood extent than the reference uniform DG2 solver in the urban banks of river Eden close to the western boundary of the domain.

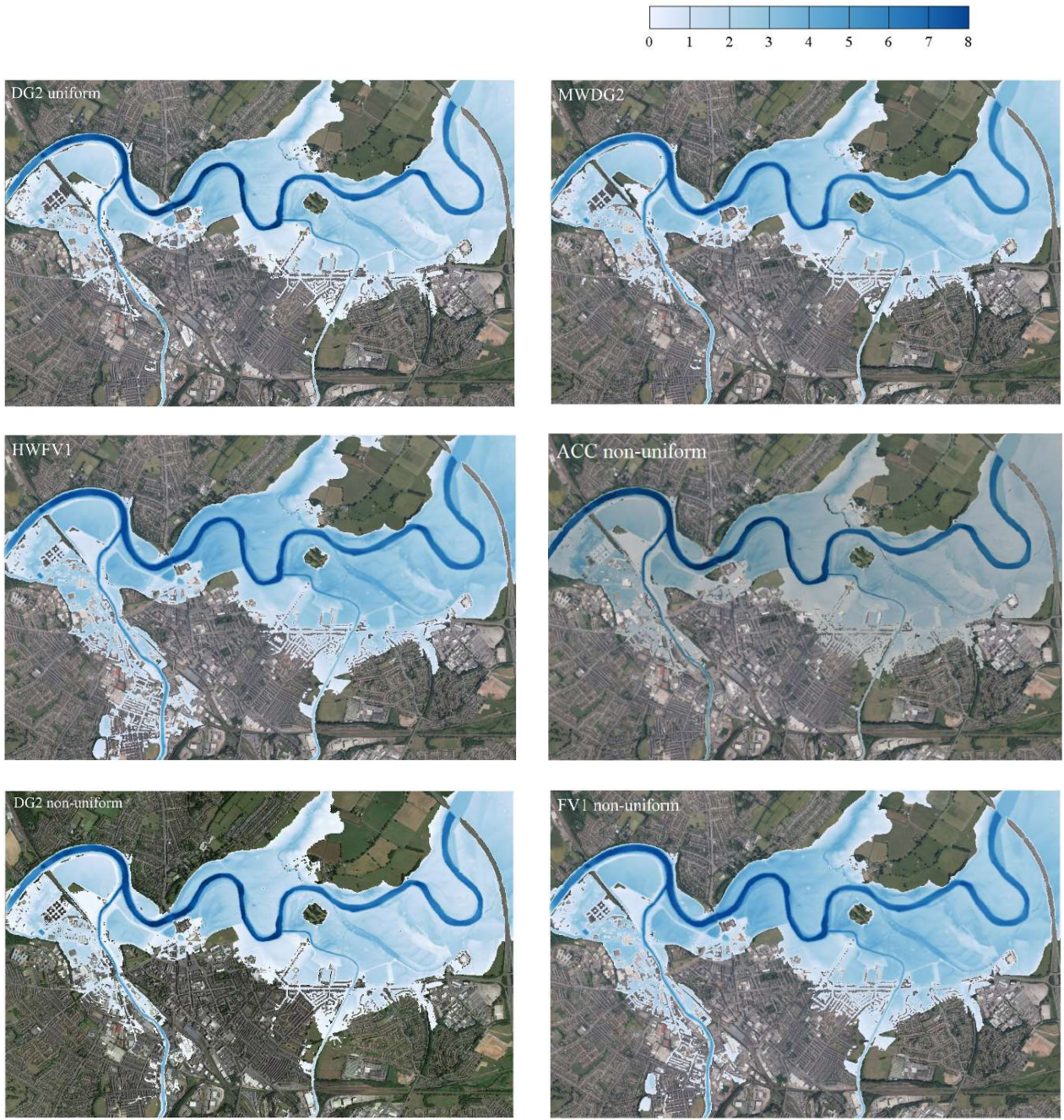

**Figure 16.** Carlisle 2005 Flooding (Section 3.3): Predicted flood inundation maps after 40 hours. Map data ©2020 Google.

**Table 6.** Carlisle 2005 Flooding (Section 3.3): hit rate (H), false alarm ratio (F) and critical success index (C) for the flood maps predicted by the adaptive MWDG2 and HWFV1 and non-uniform DG2, FV1 and ACC solvers at 10 h, against the reference predictions by the uniform DG2 solver.

| Solver | Adaptive MWDG2 | Adaptive HWFV1 | Non-uniform DG2 | Non-uniform FV1 | Non-uniform ACC |
|---|---|---|---|---|---|
| H | 0.97 | 0.99 | 0.98 | 0.99 | 0.96 |
| F | 0.005 | 0.10 | 0.008 | 0.08 | 0.05 |
| C | 0.97 | 0.89 | 0.97 | 0.90 | 0.92 |

**3.3.4 Grid coarsening ability and runtime cost**

Figure 17 shows the time evolution of the number of elements used by the adaptive MWDG2 and HWFV1 solvers during the 68-hour long simulation, along with the (constant) number of elements forming the non-uniform and uniform grids. As two distinctive stages of flooding (up to 40 hours) and recession (after 40 hours) take place, it is expected that the number of elements shows a rising and falling trend. As the initial grid predicted by the HWFV1 solver is relatively dense with only 4% fewer elements than the uniform grid (Figure 12), only a slight variation in its number of elements is observed. However, the MWDG2 solver shows a more discernible variation in the number of elements. Initially, it uses 25% fewer elements than the uniform and 14% fewer elements than the non-uniform grids. Its number of elements increases to reach its peak by 40 hours, which is still 21% fewer than the number of elements on the uniform grid and 10% fewer than the non-uniform grid. As the flood begins to recede, the coarsening of the grid leads to a gradual decrease in the number of elements, in a way that at the end of the simulation the grid only has 2% more elements than the initial grid.

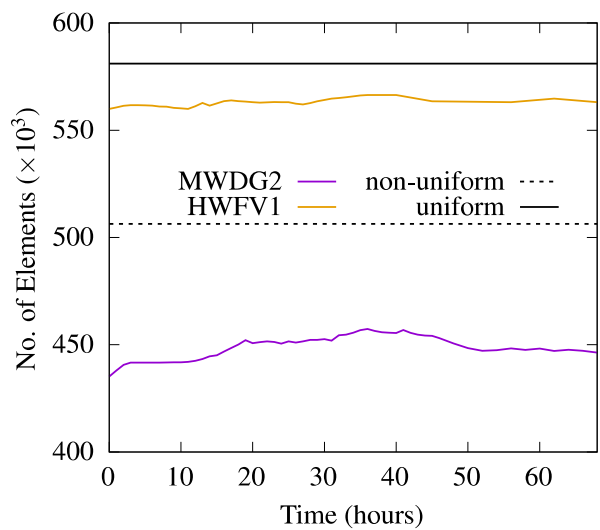

**Figure 17.** Carlisle 2005 Flooding (Section 3.3): Variation of the number of elements used by the MWDG2 and HFV1 solvers over the 68-hour long simulation. The numbers of elements used in the uniform and the MW-based non-uniform grids are marked as the solid and dotted black horizontal lines, respectively.

The CPU runtimes taken by the solvers to complete the 68-hour simulation are summarised in Table 7. The uniform DG2 yields the highest cost of 1024 hours, while the MWDG2 and HWFV1 solvers require 682 and 265 hours, making them 1.5 and 3.8 times faster than uniform DG2, respectively. Among the non-uniform solvers, as expected, the ACC is the most efficient by only taking 68 hours to finish the simulation that is 15 times faster than uniform DG2. The non-uniform FV1 and DG2 solvers also are 7.5 and 2.8 times faster than the uniform DG2 by taking 136 and 362 hours.

Table 7. Carlisle 2005 Flooding (Section 3.3): Solver runtimes for Uniform DG2, Adaptive MWDG2 and HWFV1, and non-uniform DG2, FV1 and ACC solvers, over the 10 h long simulation.

| Solver | Uniform DG2 | Adaptive MWDG2 | Adaptive HWFV1 | Non-uniform DG2 | Non-uniform FV1 | Non-uniform ACC |
|---|---|---|---|---|---|---|
| Runtime | 1024 hours | 682 hours | 265 hours | 362 hours | 136 hours | 68 hours |

For this case study, choices amongst the solvers can be established for modelling real-world fluvial floods where the realistic urban-rural terrains increase the complexity of the flooding processes (i.e., riverine flow, channel hydraulics and inundation). It is evident that the MWDG2 and non-uniform DG2 solvers are useful for leveraging the piecewise-planar representation of the topography data and the flow, both in terms of selectively adapting the grid to the mix of urban/rural terrain shapes and capturing flood flow dynamics from multiple sources. Non-uniform DG2 on the graded grid generated by the MRA of MWs allows avoiding the extra cost of dynamic adaptation, especially for the scenarios with the dominance of rural floodplains. Non-uniform ACC is shown to be less accurate than the DG2-based solvers in modelling the complex flood flow dynamics. However, when simulating fluvial flooding over complex terrain driven by multiple inflows, non-uniform ACC has led to closer results to that of non-uniform DG2 than non-uniform FV1 and HWFV1 and faster runs, making it a more favourable choice modelling large-scale complex flooding scenarios.

## 4. Summary and conclusions

This paper proposed new static non-uniform grid solvers, formed by adapting raster-based second-order discontinuous Galerkin (DG2), first-order finite volume (FV1) and ACC solvers on an optimised non-uniform grid generated using the scalability of multiresolution analysis (MRA) of multiwavelets (MWs). These solvers were evaluated alongside dynamically adaptive MWDG2 and HWFV1 solvers to reproduce three large-scale flooding case studies with increasing level of complexity. The first test simulated an industry-standard flash flood inside a valley with rugged terrain that generated a mainstream flow with both rapidly and gradually propagation stages. The second test simulated a hypothetical defence failure that resulted in a gradually propagating flood spreading over a low-lying area with complex topography. The last test case reproduced a real-world fluvial flood from multiple riverine inflows over a mixed urban/rural area with realistic fine resolution topography. The water depth/velocity hydrographs and the inundation maps obtained from the adaptive and non-uniform grid solvers were compared to reference predictions from the raster-based DG2 solver. The hydrographs were further analysed based on root-mean-square error (RMSE) while the inundation maps were quantitatively compared based on hit rate, false alarm and critical success index metrics. The computational performance of the solvers was also assessed based on the respective runtimes taken to finish the simulations.

The results suggested that when modelling rapidly propagating floods, the adaptive MWDG2 and HWFV1 solvers are cheaper alternatives to retrieve the modelling quality of the expensive uniform DG2 solver. In particular, the MWDG2 solver could more accurately capture velocity transients than the adaptive HWFV1 solver due to its smooth the piecewise-planar representation of the flow and topography data. However, when modelling gradually propagating floods that are mostly driven the features of the topography, the overhead cost of the MRA in adaptive MWDG2 or HWFV1 solvers may lead to unnecessary increase in the runtime cost. For these types of flows the non-uniform DG2, FV1 and ACC solvers can be more efficient alternatives. Amongst them, non-uniform FV1 is the least favourable as it tended to overpredicts the flood extent and intensity, due to its local first-order piecewise-constant accuracy in representing both flow variables and topography. Non-uniform ACC delivered better predictive quality than non-uniform FV1, due to its second-order spatial accuracy in representing the discharges on staggered grid stencil. Moreover, non-uniform ACC led to predictions that are comparable to those from the uniform DG2 solver but at a much lower computational cost, making it the most efficient solver amongst the others when the flood flow is in the lower range of subcritical regime. For rapidly propagating flows, the non-uniform DG2 could be vulnerable to slight velocity noises caused by its approximate evaluations of the solution limits across non-homogeneous interfaces, but these noises can be avoided by using the adaptive MWDG2 solver to model this type of flows. However, for gradually propagating flows, the non-uniform DG2 is a more accurate and efficient alterative to MWDG2 that also outperforms the other non-uniform solvers in capturing the flow transients and small-scale velocity variations occurring form the transitions between urban and rural topographies. The superiority of non-uniform DG2 is expected given its smooth piecewise-planar representation and evolution of the flow data in a much more complex formulation than ACC and FV1, that makes its runtime cost 5-6 times more than the ACC and 3-4 times more than the FV1, but still at least twice more efficient than a uniform DG2 simulation. Parallel processing is therefore a likely way forward to boost the efficiency of non-uniform DG2 for fluvial and pluvial flood simulations over urban catchments requiring a modelling with a range of resolution scales.

**Data Availability Statement**

The Fortran 2003 and C++ implementations of the static non-uniform grid solvers are available upon request by email for scientific collaboration. The Fortran 2003 implementation of the dynamically adaptive solvers is available from Zenodo (Sharifian & Kesserwani, 2020), with instructions for running them provided in Kesserwani and Sharifian (2020). The simulation results for the case studies in Section 3 are available from Zenodo (Sharifian & Kesserwani, 2021). Due to access restrictions, readers are invited to contact the Environment Agency for access to the DEMs used in case studies in Sections 3.

**Acknowledgements**

This work was supported by the UK Engineering and Physical Sciences Research Council (EPSRC) grant EP/R007349/1. It is part of the SEAMLESS-WAVE (SoftwarE infrAstructure for Multi-purpose fLood modElling at various scaleS based on WAVElets) project. For information about the project visit https://www.seamlesswave.com.